\documentclass[11pt]{article} % use larger type; default would be 10pt
\usepackage{apacite}
\usepackage[square]{natbib}

\usepackage[utf8]{inputenc} % set input encoding (not needed with XeLaTeX)
\usepackage[toc, page]{appendix}
\usepackage{amsfonts}
\usepackage{bm}
\usepackage{amsmath}
\usepackage{bbm}
\numberwithin{equation}{section}
\numberwithin{figure}{section}
\numberwithin{table}{section}

\usepackage[ruled,vlined, linesnumbered]{algorithm2e}

\SetKwInput{KwInput}{Input}                % Set the Input
\SetKwInput{KwOutput}{Output} 

\usepackage[total={6.5in, 8in}]{geometry} % to change the page dimensions 
\usepackage{graphicx} % support the \includegraphics command and options
\usepackage{float}
\usepackage{caption}
\usepackage{subcaption}
\usepackage{rotating}

\usepackage{tabularx}
\usepackage{booktabs} % for much better looking tables
\usepackage{array} % for better arrays (eg matrices) in maths
\usepackage{paralist} % very flexible & customisable lists (eg. enumerate/itemize, etc.)
\usepackage{verbatim} % adds environment for commenting out blocks of text & for better verbatim
\usepackage{hyperref}
\hypersetup{pdftex,colorlinks=true,allcolors=blue}
\usepackage{hypcap}
\usepackage{amsthm}
\usepackage{fancyhdr} % This should be set AFTER setting up the page geometry
\usepackage{sectsty}
\allsectionsfont{\sffamily\mdseries\upshape} % (See the fntguide.pdf for font help)
\usepackage[nottoc,notlof,notlot]{tocbibind} % Put the bibliography in the ToC
\usepackage[titles,subfigure]{tocloft} % Alter the style of the Table of Contents

\usepackage[font=small,labelfont=bf,tableposition=top]{caption}

\usepackage{authblk}
\usepackage[export]{adjustbox}
\usepackage{mathtools}

\theoremstyle{definition}

\newcommand{\PreserveBackslash}[1]{\let\temp=\\#1\let\\=\temp}
\newcolumntype{C}[1]{>{\PreserveBackslash\centering}p{#1}}
\newcolumntype{R}[1]{>{\PreserveBackslash\raggedleft}p{#1}}
\newcolumntype{L}[1]{>{\PreserveBackslash\raggedright}p{#1}}
\newcolumntype{P}[1]{>{\centering\arraybackslash}p{#1}}

\DeclareCaptionLabelFormat{andtable}{#1~#2  \&  \tablename~\thetable}

\providecommand{\keywords}[1]
{
  \small	 
  \textbf{\textit{Keywords---}} #1
}

\title{Towards systematic intraday news screening: a liquidity-focused approach\thanks{
    This work benefits from the financial support of the Chaires Machine Learning \& Systematic Methods. The author would like to thank  Qinkai Chen and Quentin Jacob for very useful comments.
  }}

\author{Mathieu Rosenbaum $^1$ \\ \vspace{-1em} mathieu.rosenbaum@polytechnique.edu
\and
Jianfei Zhang $^{1,2}$ \\ \vspace{-1em} jianfei.zhang@polytechnique.edu}
\date{%
$^1$ \footnotesize École polytechnique, CMAP, Institut Polytechnique de Paris, 91120 Palaiseau, France \\%
$^2$ \footnotesize Exoduspoint Capital Management, 32 Boulevard Haussmann, 75009 Paris, France \\[2ex]%
\today}

\begin{document} 
    \maketitle
    \begin{abstract}
        News can convey bearish or bullish views on financial assets. Institutional investors need to evaluate automatically the implied news sentiment based on textual data. Given the huge amount of news articles published each day, most of which are neutral, we present a systematic news screening method to identify the ``true'' impactful ones, aiming for more effective development of news sentiment learning methods. Based on several liquidity-driven variables, including volatility, turnover, bid-ask spread, and book size, we associate each 5-min time bin to one of two specific liquidity modes. One represents the ``calm'' state at which the market stays for most of the time and the other, featured with relatively higher levels of volatility and trading volume, describes the regime driven by some exogenous events. Then we focus on the moments where the liquidity mode switches from the former to the latter and consider the news articles published nearby impactful. We apply naive Bayes on these filtered samples for news sentiment classification as an illustrative example. We show that the screened dataset leads to more effective feature capturing and thus superior performance on short-term asset return prediction compared to the original dataset.     
    \end{abstract}

    \section{Introduction}
        The price of a financial asset is driven by endogenous activities, such as self-reflexive trades, and also exogenous information. A main
        component of the latter comes from news releases. Nowadays, the financial market is becoming increasingly efficient, yet the embodiment of new information 
        transferred by news in asset price is rarely accomplished instantaneously. Quick and effective estimation of news sentiment, 
        \textit{i.e.} whether the view given by a news article is bullish or bearish, can give profitable opportunities to investors. As said in \cite{pedersen2019efficiently}, 
        ``financial markets are efficiently inefficient'', in the idea that professional investors with superior performance are compensated for their costs
        and risks, and the competition among them makes markets almost efficient. 
        Given the large number of news publications every day, manual analysis of each piece of news is infeasible. To assess efficiently the impact of a news release on the market price of its associated financial asset, institutional investors then need to develop automatic news sentiment evaluation methods. \\
        
        \noindent
        Numerous works using news data to predict financial assets' price movements exist in the literature. In \cite{chan2003stock}, it is documented that the monthly returns of the stocks
        associated with public news releases are less likely to reverse than those without identifiable news publication, suggesting that news can publish some information concerning the ``fair'' value of stocks. \cite{jiang2021pervasive}
        extend the study with intraday returns in the same spirit. A long/short trading strategy exploiting the news-driven price drifts is shown to generate abnormal profit.
        In these works, the views transferred by certain news releases on the stocks of interest, whether bullish or bearish, are identified by the associated post-news market reactions. Thus to make an investment decision, investors have to wait until the emergence of some significant price drifts after news publications. To react more quickly, sentiment evaluation methods based on textual data are required. \cite{tetlock2007giving} applies the Harvard-IV psychosocial dictionary on the articles from \textit{The Wall Street Journal} to estimate the pessimistic pressure
        on the Dow Jones Industrial Average index. \cite{loughran2011liability} construct a customized dictionary more adapted to financial text using the term frequency-inverse document frequency (tf-idf) method. Stock market sentiment lexicons based on microblogging data are developed in \cite{oliveira2016stock} and \cite{renault2017intraday} by computing several statistical measures. \cite{ke2019predicting} estimate the tone weights of sentiment-charged words via a topic modeling method, and compute the article-level
        sentiment score using a regularized version of maximum likelihood estimation. \\
        
        \noindent
        Another strand of research is using deep learning methods on news data in the same vein as current practices in natural language processing. \cite{ding2014using} transform news publications 
        into structural events using Open Information Extraction techniques. A novel neural tensor network is used in \cite{ding2015deep}
        to represent news text as dense vectors. Then the representations are put into a convolutional neural network for predicting future stock price movements. The dependence 
        among sequential news releases is considered in \cite{hu2018listening} by proposing an attention-based recurrent neural network.  \cite{chen2021stock} develops a fine-tuned bidirectional encoder representations
        from transformers, see \cite{kenton2019bert}, to produce contextualized word embeddings, which are then fed into a recurrent neural network to output news classification results. \\
          
        \noindent
        To develop a news sentiment learner, one needs firstly labeled samples, \textit{i.e.} bullish or bearish news releases, for model training. However, the sentimental nature of a news article is not explicitly marked in real life. 
        In the case of Bloomberg News\footnote{https://www.bloomberg.com/professional/product/event-driven-feeds/}, the first step of model development consists of manual labeling by human experts, in the idea to 
        isolate ``true'' sentiment contained in news from realized price actions. However, this method can be exposed to subjective biases. More importantly, one has to repeat the manual labeling when moving to a new training set, which can be costly. More efficient systematic news labeling methods are preferred in this case. 
        A common practice in the research community is to take the sign of the share price movement following a news publication as the ground truth label \citep{chen2021stock, ding2014using,ding2015deep, hu2018listening}. This is understandable since news sentiments are expected to predict positively the price drifts of associated assets. However, not every news article conveys a directional view of the underlying asset price. Labeling news only based on post-publication returns would result in a non-negligible proportion of falsely labeled samples in the training set, which causes extra uncertainties for model learning. \cite{chen2021stock} keeps only a small portion of samples with extreme market realized returns for model training to reduce the impact of neutral news. Still, we believe that relying solely on the posterior return is incomplete, given the low signal-to-noise ratio of return data. \\

        \begin{figure}[htbp]
            \centering
            \begin{subfigure}{.5\textwidth}
                \centering
                \includegraphics[width=\linewidth]{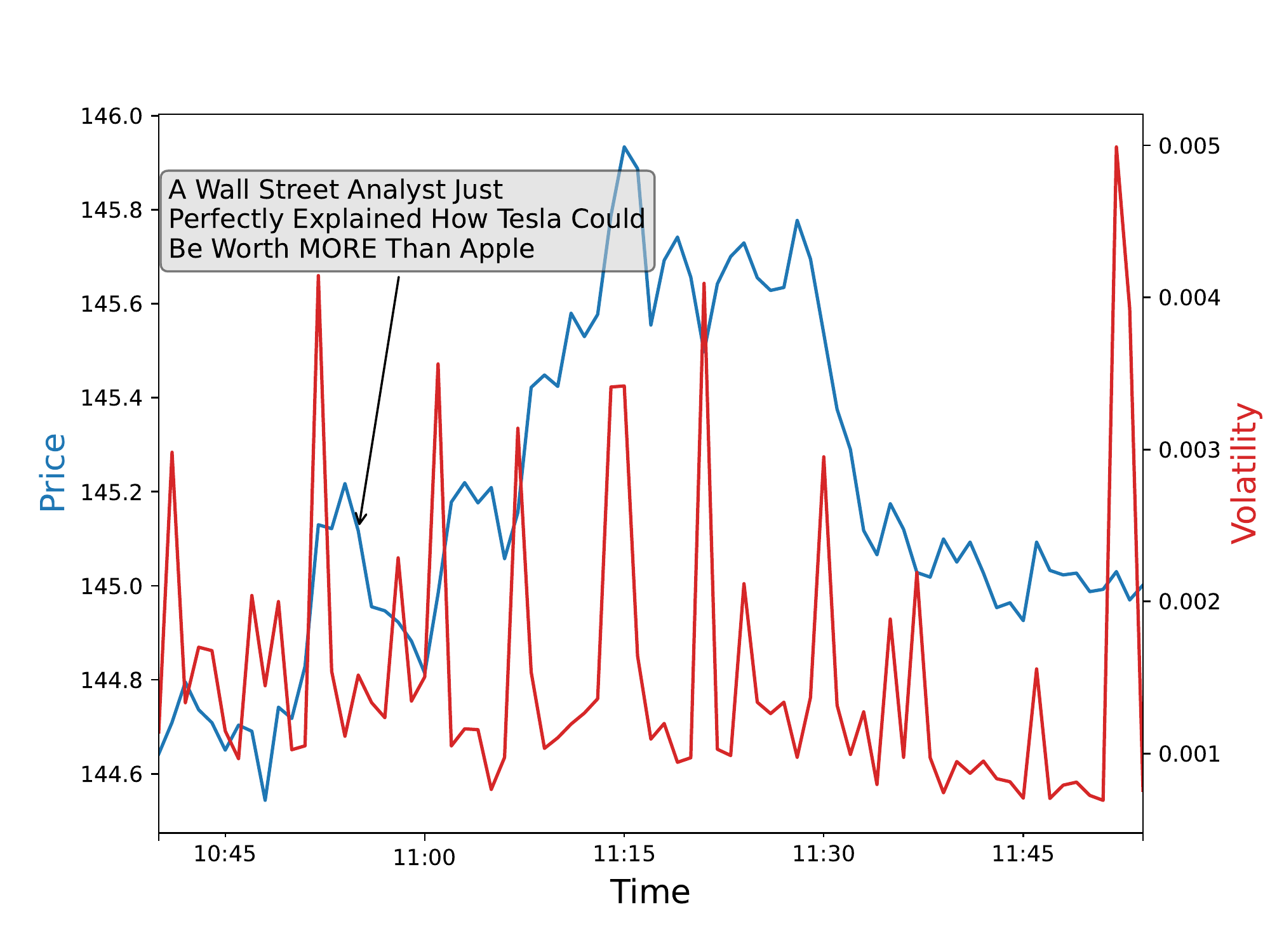}
            \end{subfigure}%
            \begin{subfigure}{.5\textwidth}
                \centering
                \includegraphics[width=\linewidth]{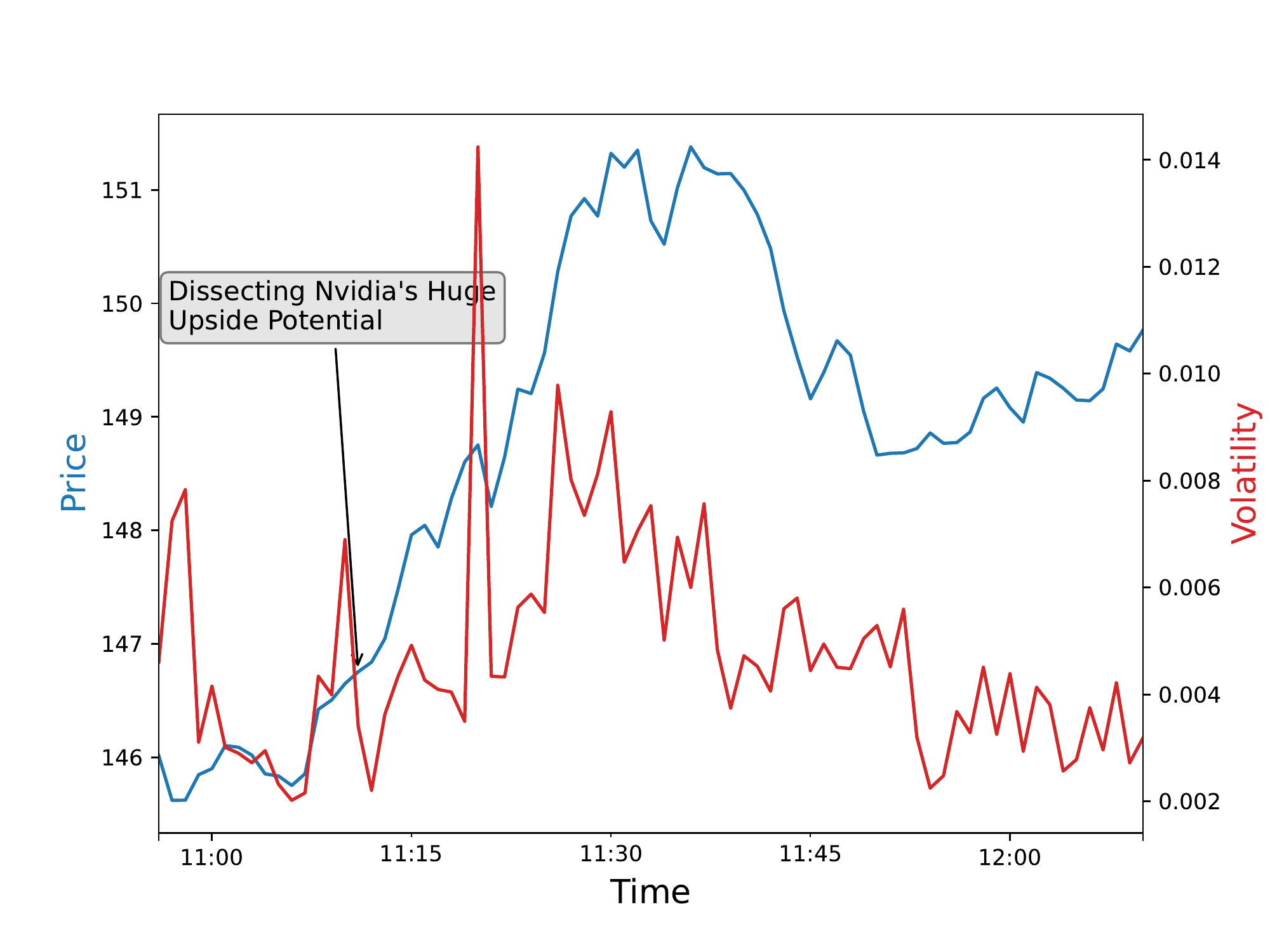}
            \end{subfigure}
            \caption{Dynamics of stock price and volatility of Apple(left) and Nvidia(right) around two news publications, indicated by the arrows.}
            \label{fig:news_plot}
        \end{figure}

        \noindent
        In this paper, we aim at identifying the ``true'' impactful news releases, \textit{i.e.} those imply positive or negative views on the associated assets, with a particular focus on the liquidity states of the assets of interest. Figure \ref{fig:news_plot} illustrates our main motivation by examining stocks' price and volatility dynamics around 
        news publications with two particular examples. Both news releases would be considered positive if we look at only the short-term realized return. Yet the first one is more 
        likely to be a piece of neutral news for Apple, which is consistent with the mediocre volatility fluctuation. As for the case of Nvidia, we observe a more significant volatility change when the message 
        conveys a positive view on the company's potential. Note that here we apply the model with uncertainty zones introduced by \cite{robert2011new, robert2012volatility} for effective computation of
        high-frequency volatility. Thus, we can conceive a two-step news screening approach. First, we divide the daily core trading session into multiple nonoverlapping time bins of equal size, and for each bin, we estimate the realized price volatility. These estimations can be sorted out using clustering methods into two clusters representing two distinct liquidity modes. We designate the one with a lower volatility level as \textit{Mode 1} and the other as \textit{Mode 2}. Second, we focus on the jumps from \textit{Mode 1} to \textit{Mode 2}, and consider the news articles published around these jumps impactful. That is, these jumps are thought to have caused these volatility changes. \\
        
        \noindent
        The above method is consistent with the findings in \cite{gross2011machines}. Based on news data concerning 40 stocks actively traded at the London Stock Exchange, preprocessed by the Reuters NewsScope Sentiment Engine, \cite{gross2011machines} suggests that the news with high relevance impact significantly volatilities and trading volumes.
        To get a more robust liquidity mode fitting, the actual indicator set chosen in this work includes four common liquidity-driven variables, \textit{i.e.} volatility, turnover, bid-ask spread, and book size \citep{binkowski2022endogenous}. Instead of using classical clustering methods like K-means to distinguish two liquidity modes, we apply the jump model introduced in \cite{bemporad2018fitting} to take also the ordering of observations into account. In the jump model, we can easily penalize frequent mode switches, and thus some mode persistence is favored. This is relevant for time series describing intraday liquidity conditions. \\
        
        \noindent
        \cite{joulin2008stock, marcaccioli2022exogenous} investigate the volatility fluctuations around the price jumps induced by news releases, suggesting that they exhibit different dynamical patterns to those arising from endogenous activities. One can thus monitor the volatility dynamics around each news publication, and then decide whether the release has impacted the market according to the observed features. However, we have to consider the following inconveniences when applying this method in practice. First, news screening one by one is very time-consuming given the large size of the news dataset. Second, when some other liquidity-driven variables in addition to volatility are considered, as in the case of our current work, the dynamical features of these variables differentiating between the exogenous and endogenous events need to be specified explicitly, and then news classification rules should be modified accordingly. \\

        \noindent
        Our approach is very efficient by structuring the task into two steps: liquidity mode fitting on a time-bin basis and associating news releases with detected liquidity mode jumps. The fitting results can also be used for news data from other providers or any other exogenous events/signals in a similar manner. The jump model fits data in a nonparametric way, allowing us to be agnostic about the dynamical properties of each measured variable. Other variables can be easily tested in the same manner. 
        Note that only the news releases that happened during the daily core trading sessions are concerned by our screening approach.
        Once the impactful releases are targeted, we mark them as positive or negative by the signs of the post-news returns of the associated assets, similarly to \cite{chen2021stock, ding2014using,ding2015deep, hu2018listening}.    
        Then various supervised learning methods can be applied to the labeled samples to learn sentiment-charged features. In this work, we focus on the effect of our news screening method instead of developments of alternative sentiment learning methods. To illustrate the idea, we fit two Bernoulli naive Bayes classifiers (NBCs) respectively on the original intraday news data and the dataset filtered by liquidity mode changes. Based on out-of-sample tests, the classification results given by the later classifier are more consistent with the post-news asset price movements than the former, \textit{i.e.} in average the news articles with high probability to be positive (negative) assigned by the latter classifier show more significant and persistent positive (negative) post-news price drift than the ones sorted out by the former classifier. This indicates that the screened dataset includes fewer falsely labeled samples, and thus can lead to more effective sentiment learning. \\

        \noindent
        The paper is organized as follows. In Section \ref{sec:data}, we describe firstly the data involved in this study and related preprocessing procedures. The application of the jump model on liquidity mode fitting is detailed in Section \ref{sec:mode_fit}. Numerical results with market data are presented. We then present how to use the fitted mode sequences to identify impactful news releases in Section \ref{sec:news_learn}. The relevance of our method for more effective news sentiment learning is then illustrated through numerical experiments. Finally, we conclude with our main findings in Section \ref{sec:conc}.

    \section{Data and preprocessing}
    \label{sec:data}
        In this work, we use the Daily Trade and Quote (TAQ) dataset\footnote{https://www.nyse.com/market-data/historical/daily-taq} for computing the considered liquidity-driven
        variables. The TAQ dataset covers all stocks traded in the US market. To avoid any results biased by small-cap names, we focus on the components of S\&P 500. 
        The following variables are measured on consecutive bins of five minutes during the daily core trading session:
        \begin{itemize}
            \item \textit{Average bid-ask spread in ticks} $(\phi)$. For each second, we record the last observed bid-ask spread. The average value for each 5-minute bin is computed and its ratio against the tick
                   size is kept. 
            \item  \textit{Traded value/turnover} $(V)$, is the total value traded during each 5-minute bin.
            \item \textit{Volatility} $(\sigma)$. Instead of using the Garman-Klass volatility as in \cite{binkowski2022endogenous}, here we take the one based on the model with uncertainty zones, whose 
                        effectiveness on high-frequency data is shown in \citep{robert2011new,robert2012volatility}.
            \item  \textit{Average book size} $(B)$. We record the average volume available at the best bid and ask prices of each second inside each 5-minute bin.
        \end{itemize}
        \noindent
        After removing the data concerning the first and last 15 minutes of the daily core trading period\footnote{We recall that the regular trading hours for the US stock market are 9:30 a.m. to 4:00 p.m.}, for each (stock, day) pair, we get a time series of length $T=72$. Intraday seasonalities of liquidity variables,  induced by certain fundamental reasons, are well known \citep{binkowski2022endogenous, lehalle2018market},
        and our objective is to identify the short-term impacts of news beyond these effects. For each observation $o_t^{S,d}\in\{\phi^{S,d}_t, V^{S,d}_t, \sigma^{S,d}_t, B^{S,d}_t\}$ 
        of stock $S$ on day $d$ with $t=1, \dots, T$, we use the following stationarization procedure:
        $$
            o^{S,d}_t = \frac{\log o^{S,d}_t - \frac{1}{D}\sum_{d^\prime=1}\log o^{S,d^\prime}_t}{\text{IQR}(\log o^{S, 1}_t, \dots, \log o^{S, D}_t)} \, ,
        $$
        where $\text{IQR}(\cdot)$ means the difference between the 75th and 25th percentiles of the data, and $D$ represents the total available days on the dataset under study. 
        We use $\text{IQR}$ instead of the ordinary standard deviation to reduce the impact of outliers. In this paper, data covering 2017-01-01 $\sim$ 2019-12-31 is selected for examining
        the association between intraday liquidities and news data, \textit{i.e.} $D\simeq 750$. After this preprocessing, the intraday seasonalities can be mostly removed 
        and all the variables of all stocks have similar scales, which is essential for the fitting of intraday liquidity states, as will be detailed in the following. \\

        \noindent
        Table \ref{tab:bloom} gives several samples of the Bloomberg News data that we used in this work. Note that we evaluate the news sentiment only based on the headlines.
        The \textit{Score} and \textit{Confidence} fields give sentiment estimation of the news articles, which are based on Bloomberg's proprietary classification algorithm.   
        The \textit{Score} says whether a piece of news is bullish(1), bearish(-1), or neutral(0), with the \textit{Confidence} indicating the reliability of this classification. We can thus compute an 
        \textit{composed score} $:=$ \textit{score} $\times$ \textit{confidence}. 
        Naturally, we expect on average the news releases with the highest \textit{composed score} are followed with significantly positive price drift and the opposite for the ones with the lowest \textit{composed score}. The predictive power of this score will be compared with the sentiment score given by the NBCs, as will be shown later. \\
        \begin{table}[hbtp]
            \begin{center}
              \begin{tabular}{P{16em}P{8em}ccc}
                \toprule
                \textbf{Headline} & \textbf{TimeStamp} & \textbf{Ticker} & \textbf{Score} & \textbf{Confidence} \\
                \midrule
                1st Source Corp: 06/20/2015 - 1st Source announces the promotion of Kim Richardson in St. Joseph & 2015-06-20T05:02:04.063 & SRCE & -1 & 39 \\
                \midrule
                Siasat Daily: Microsoft continues rebranding of Nokia Priority stores in India opens one in Chennai & 2015-06-20T05:14:01.096 & MSFT & 1 & 98 \\
                \midrule
                Rosneft, Eurochem to cooperate on monetization at east urengoy & 2015-06-20T08:01:53.625 & ROSN RM & 0 & 98 \\
                \bottomrule
              \end{tabular}
          \end{center}
          \caption{Several samples of the Bloomberg News data.}
          \label{tab:bloom}
        \end{table}%
        
\begin{figure}[htbp]
            \centering
            \includegraphics[width=\linewidth]{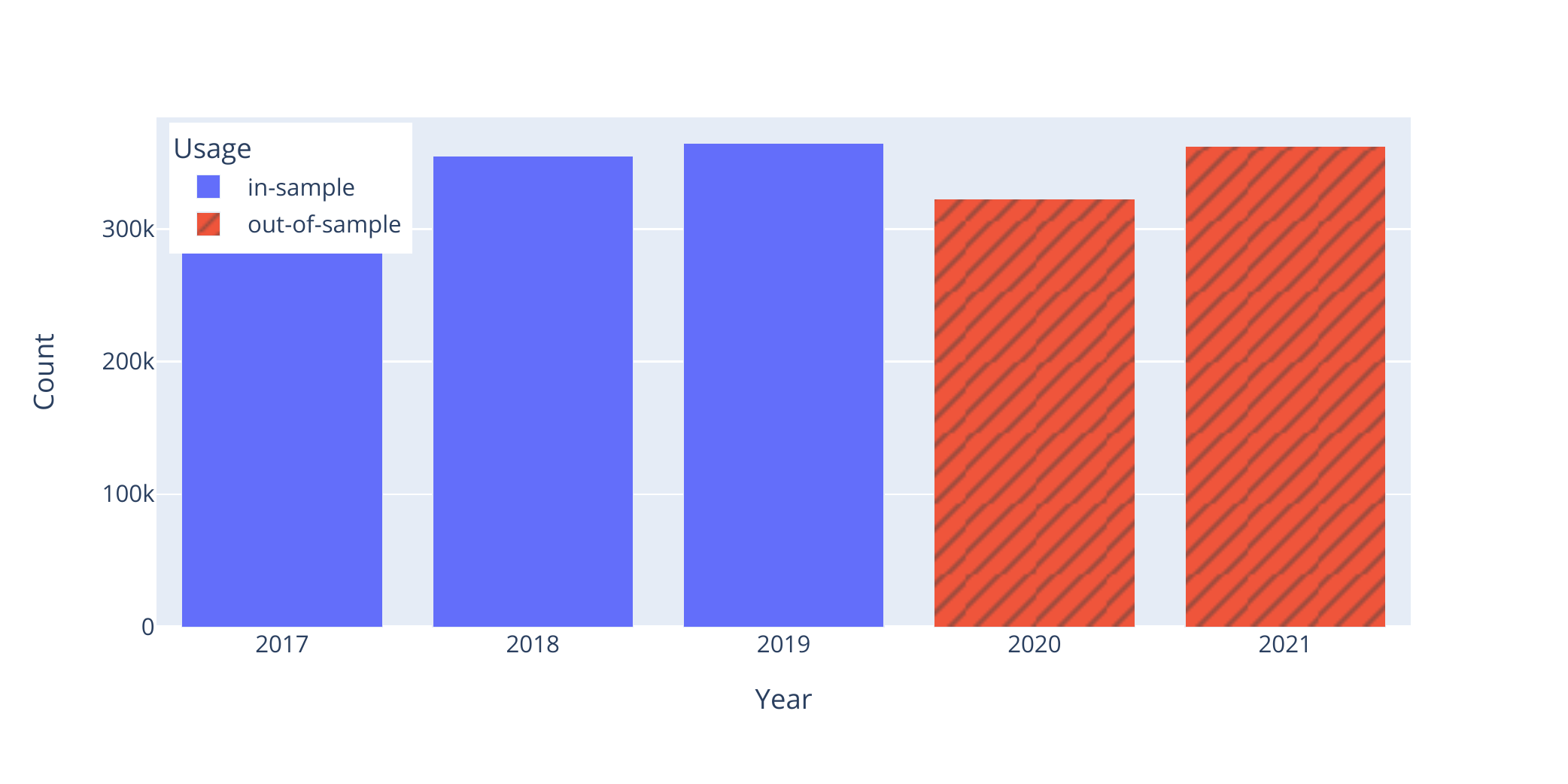}
            \caption{Count of the news releases concerned in current work.}
            \label{fig:news_dist}
        \end{figure}

        \noindent 
        Figure \ref{fig:news_dist} shows the number of news releases related to our work. We screen the news headlines published during the years 2017$\sim$ 2019 based on the liquidity-driven variables as introduced above. 
        Thus only the ones released during the daily trading hours is concerned for this step. 
        We then fit a news classifier to output sentiment scores capturing the sentiment-charged features based on the samples picked out.
        The classifier is evaluated on all the intraday news headlines published during 2020-01-01 $\sim$ 2021-12-31, with a particular focus on the predictive power of the sentiment scores on the short-term post-publication price drifts. 

    \section{Liquidity mode fitting}
        \label{sec:mode_fit}
        \subsection{Jump model}
            Given a sequence of data pairs $(x_t, y_t)_{t=1,\dots, T}$ with $x_t\in\mathcal{X}, y_t\in\mathcal{Y}$ and the number of modes $K$, the jump model introduced in \cite{bemporad2018fitting}
            outputs a mode sequence $(m_t)_{t=0,\dots, T}$ with $m_t\in\{1, \dots, K\}:=\mathcal{K}$, and the parameter $\theta_m\in \mathbb{R}^a$ associated with each $m\in\mathcal{K}$, with $a$ some positive constant. 
            The obtained $\Theta:=(\theta_1,\dots,\theta_K)$ and $M:=(m_0,\dots,m_T)$ minimize the following objective function
            $$
                J(X, Y, \Theta, M) = \sum_{t=1}^T l(x_t, y_t, \theta_{m_t}) + \sum_{k=1}^K r(\theta_k) + \mathcal{L}(M) \, ,
            $$
            where $X:=(x_1,\dots, x_T)$, $Y:=(y_1,\dots, y_T)$, $J(\cdot)$, $l(\cdot)$, $r(\cdot)$ and $\mathcal{L}(\cdot)$
            are all functions taking value on $\mathbb{R}$. The first two parts to the right of the equal sign
            represent respectively the total fitting loss of the given data and a regularization term on the model parameters. For example, when $K=1$, $l(x, y, \theta)= \|y-\theta^\prime x\|^2_2$ and 
            $r(\theta)=\lambda\|\theta\|_2^2$ with $\lambda>0$, we get the standard setting for Ridge regression. The particularity of the jump model relies on the introduction of the 
            loss $\mathcal{L}(M)$ taking the ordering of the mode sequence into account. It is defined as
            $$
                \mathcal{L}(M) = \mathcal{L}^{init}(m_0) + \sum_{t=1}^T\mathcal{L}^{mode}(m_t) + \sum_{t=1}^T\mathcal{L}^{trans}(m_t, m_{t-1}) \, ,
            $$ 
            where the loss is decomposed into three parts, the penalization on the initialization $\mathcal{L}^{init}$, the one linked to the fitting result of each timestamp $\mathcal{L}^{mode}$, and the cost concerning mode transitions $\mathcal{L}^{trans}$. As suggested in \cite{bemporad2018fitting}, this definition generalizes popular models such as hidden Markov models. Various cost functions 
            can be chosen to meet the needs of different applications. We refer to \cite{bemporad2018fitting} for more details. \\
            
            \noindent
            In our case, we apply the jump model in an unsupervised learning manner. For each stock-day pair $(S, d)$ with $S=1,\dots, N$ and $d=1,\dots, D$, we have a preprocessed 
            sequence $X^{S,d}:=(x_t^{S,d})_{t=1,\dots,T}$ where $x_t^{S,d} := (\phi^{S, d}_t, V^{S,d}_t, \sigma^{S,d}_t, B^{S,d}_t) \in \mathbb{R}^4$. The model consists in finding a mode 
            sequence $M^{S,d}:=(m^{S,d}_t)_{t=1,\dots, T} \in \mathcal{K}^T$, under which the observations associated with the same mode are more similar to each other than to those linked with other modes. We denote the $K$ mode represents by $\Theta^{S,d}:=(\theta^{S,d}_k)_{k=1,\dots,K}\in (R^4)^K$. 
            We have no prior knowledge about the initial mode $m^{S,d}_0$, and impose no mode-specific
            cost, that is, $\mathcal{L}^{init}=\mathcal{L}^{mode}=0$. We penalize frequent mode switches, expecting some degree of persistence for the fitted mode sequence.  
            Particularly, it leads to the same type of loss function as in \cite{nystrup2021feature}, which reads
            \begin{equation}
                J(X^{S,d}, \Theta^{S,d}, M^{S,d}) = \sum_{t=1}^Tl(x_t^{S,d}, \theta^{S,d}_{m^{S,d}_t}) + \lambda\sum_{t=1}^{T-1}\mathbbm{1}_{m_t^{S,d} \neq m_{t+1}^{S,d}} \, ,
                \label{eq:single_fit_obj} 
            \end{equation}
            where $l(x, \theta)=\|x-\theta\|_2^2$, with $\|\dot\|_2$ representing the $L^2$ norm, and $\lambda$ is a hyperparameter trading off between clustering the given data and mode persistence. 
            In practice, it can be chosen via cross-validation. Note that when $\lambda=0$, we will get the classical K-means solution. As for the number of modes, larger $K$ can 
            brings better fit, while it becomes less evident to interpret and involves a higher risk of overfitting. Since we focus on the market liquidity regime 
            switch caused by exogenous information, we simply take $K=2$ in the following tests. We will see in the following that the resulting two modes are separated in terms of volatility level. We let the mode associated with lower volatility be \textit{Mode 1}, and the other be \textit{Mode 2}. \\
            
            \noindent
            Therefore, the results of minimization of (\ref{eq:single_fit_obj}) are a sequence of liquidity modes associated with each time bin and two vectors of dimension four. Note that the latter are obtained based on $T=72$ observations, which still implies some potential risk of overfitting.
            To reduce this risk, we can expand the fitting on sequences of multiple days. However, as the liquidity-driven variables are not homogeneous across time, \textit{e.g.} certain periods are more volatile than the others, the fitted modes will not distribute uniformly over time. For example, we expect that \textit{Mode 2} will concentrate on periods associated with relatively larger volatility. Thus in this case, the observed mode switches are mostly the results of some market-wide events such as monetary policy announcements, instead of stock-specific news releases. In this work, we expand the fitting space in the dimension of the asset. More precisely, on day $d$ the model is fitted on the pooled set of sequences $X^d := \{(x^{S,d}_t)_{t=1,\dots,T}\}_{S=1,\dots, N}$. As market-wide activities impact all the stocks to a similar extent, and are unlikely to change abruptly at the intraday scale, the observations $(x^{S,d}_t)_{S=1,\cdots,N, t=1,\cdots,T}$ are influenced uniformly by these market-wide events across $S$ and $t$. In this way, the fitting results can reflect the short-term liquidity fluctuations from the base level related to the market environment.  Significant liquidity changes are thought to be induced by some exogenous events, \textit{i.e.} news releases in our case. We ignore the superscript $d$ for ease of notation in the following.
            The actual fitting objective reads thus
            \begin{equation}
                \arg\min_{\Theta,M}J(X, \Theta, M) = \sum_{S=1}^N\big(\sum_{t=1}^Tl(x_t^{S}, \theta_{m^{S}_t}) + \lambda\sum_{t=1}^{T-1}\mathbbm{1}_{m_t^{S}\neq m_{t+1}^{S} } \big) \, ,
                \label{eq:fit_obj}
            \end{equation}
            where $\Theta:=(\theta_1,\dots, \theta_K)$ and $M:=\{(m^{S}_t)_{t=1,\dots,T}\}_{S=1,\dots, N}$. Problem (\ref{eq:fit_obj}) can be solved with a simple coordinate-descent 
            optimization algorithm that alternates minimization with respect to $\Theta$ and $M$, which is detailed in Appendix \ref{app:jump_algo}.

        \subsection{Fitting results}
            In this part, we give several statistics on the intraday liquidity mode fitting results for the period 2017-01-01 $\sim$ 2019-12-31. We recall that in this study $K=2$ and we test several different $\lambda$ to see its effect. 
            Given the set of all fitted mode sequences $\{M^{S,d}\}_{S=1,\dots,N, d=1,\dots, D}$, we count respectively the occurrences of \textit{Mode 1} and \textit{Mode 2}. We also estimate a $2\times 2$ implicit transition matrix $\mathcal{T}$, whose entries are computed as follows:
            $$
                \mathcal{T}_{ij} = \frac{\sum_{d=1}^D\sum_{S=1}^N \sum_{t=1}^{T-1}\mathbbm{1}_{m^{S,d}_t=i, m^{S,d}_{t+1}=j}} {\sum_{d=1}^D\sum_{S=1}^N \sum_{t=1}^{T-1}\mathbbm{1}_{m^{S,d}_t==i}} \, , \quad  i,j\in\{1, 2\} \, .
            $$
            \begin{figure}[htbp]
                \centering
                \includegraphics[width=\linewidth]{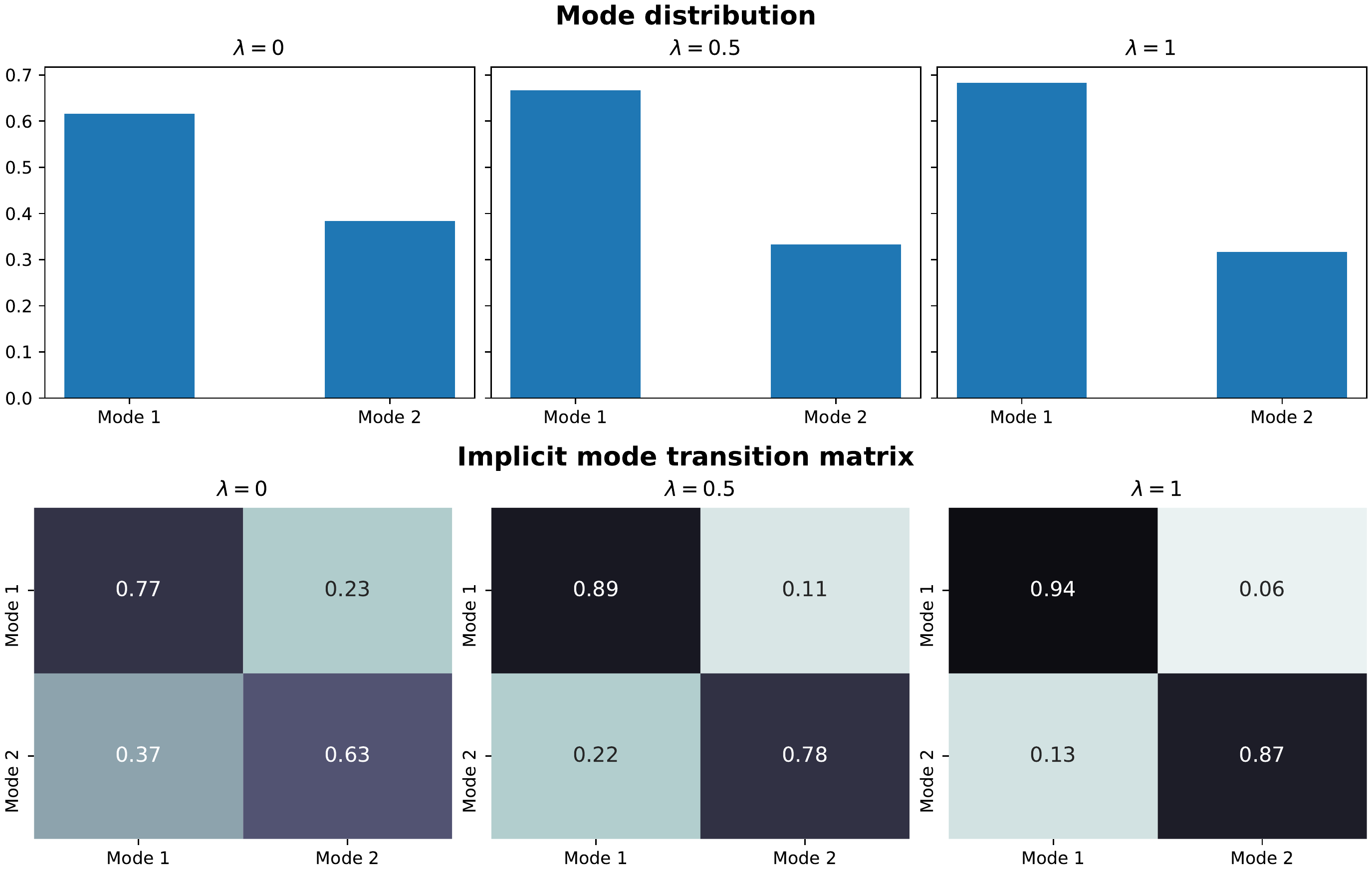}
                \caption{Mode distribution of all fitted 5-minute bins (top) and implicit mode transition matrix estimated from the fitting results.}
                \label{fig:mode_dist}
            \end{figure}

\begin{figure}[htbp]
                \centering
                \includegraphics[width=.8\textwidth]{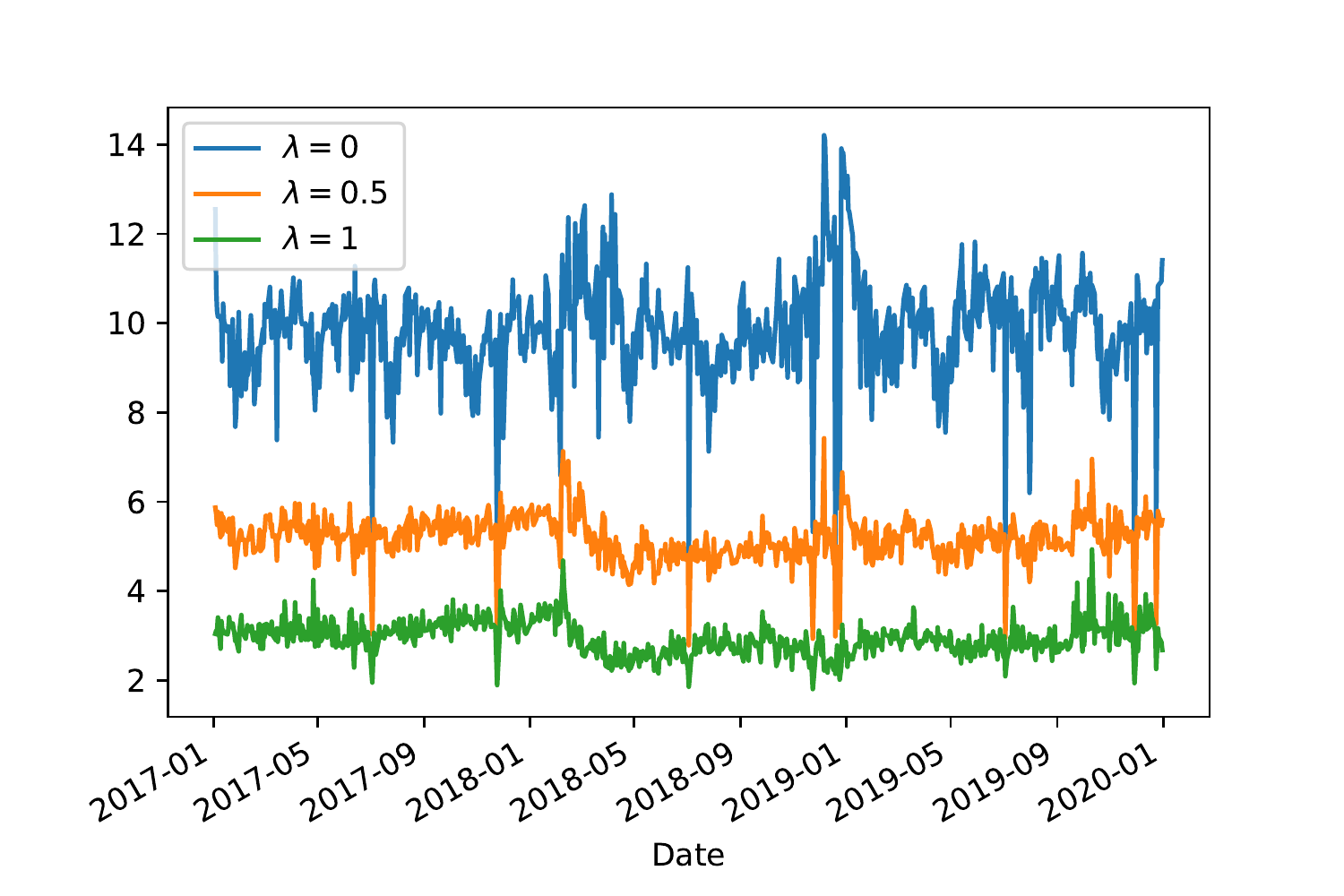}
                \caption{Average daily count of jumps from \textit{Mode 1} to \textit{Mode 2} per stock.}
                \label{fig:jump_cnt}
            \end{figure}

            \noindent
            Figure \ref{fig:mode_dist} gives the results with respect to different $\lambda$. Most of the time the market stays at \textit{Mode 1}. With larger $\lambda$, we get a slightly smaller assignment ratio for \textit{Mode 2} and more pronounced mode persistence. Note that even when we do not penalize the mode switch, \textit{i.e.} $\lambda=0$, the diagonal elements of $\mathcal{T}$ are significantly larger than the off-diagonal ones. This is consistent with the observations in \cite{binkowski2022endogenous} that all the selected liquidity-driven variables are positively autocorrelated, and thus the points of adjacent time bins are likely to be classified into the same mode. Accordingly, Figure \ref{fig:jump_cnt} gives the average daily count of liquidity mode switches from \textit{Mode 1} to \textit{Mode 2} per stock.\\
            
\begin{figure}[htbp]
                \centering
                \includegraphics[width=\linewidth]{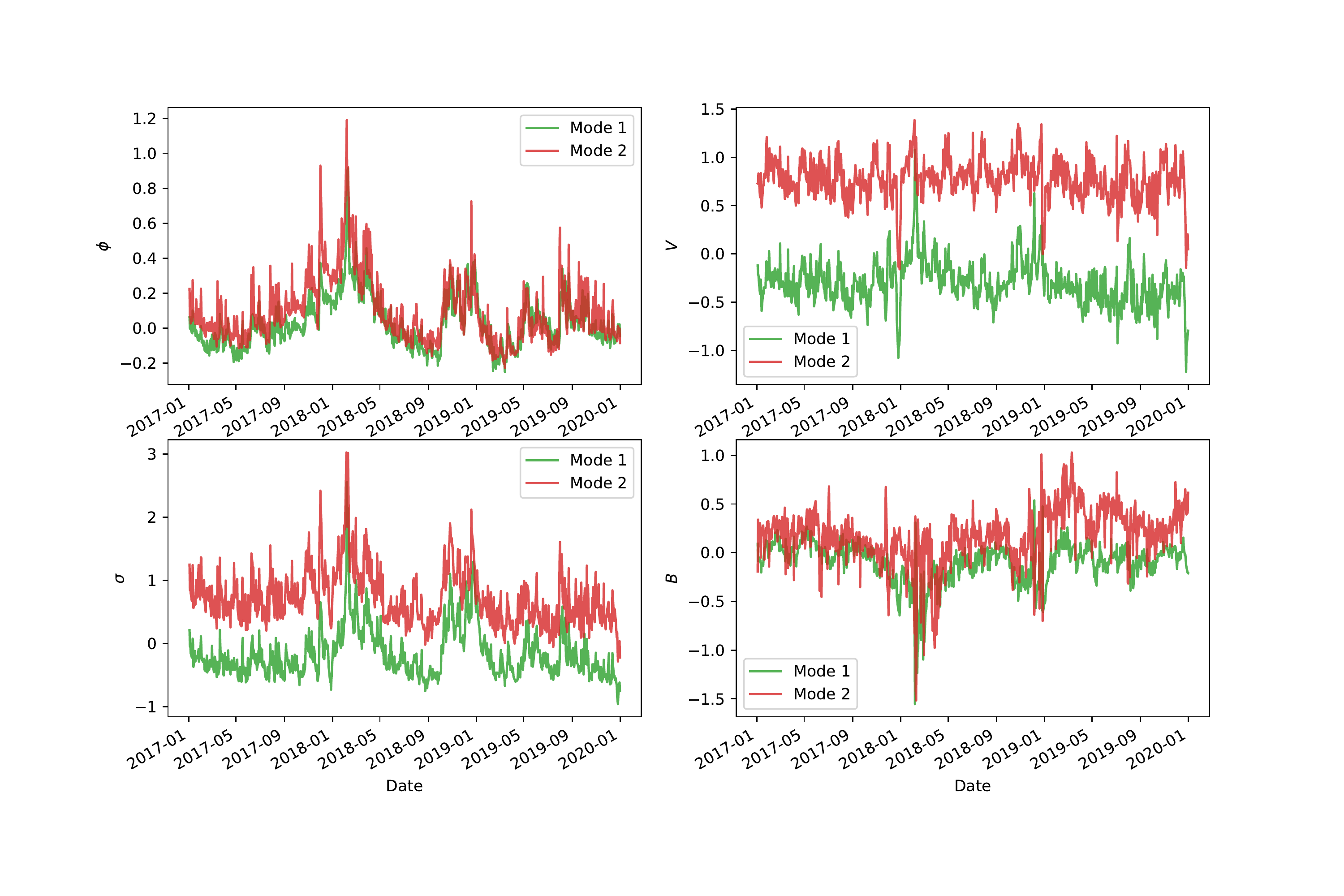}
                \caption{Historical evolution of fitted mode parameters when $\lambda=0.5$.}
                \label{fig:mode_param}
            \end{figure}

            \noindent
            Figure \ref{fig:mode_param} plots the dynamics of fitted mode parameters during our testing period when $\lambda=0.5$. As expected in Introduction, the volatility levels of the two modes are well distinct. Interestingly, it is also the case for the traded volume $V$. The differentiation of $\phi$ and $B$ between the two modes are relatively less noticeable. It is unlikely that the detected \textit{Mode 2} corresponds mostly to the time slots with endogenous volatility spikes, which are usually accompanied by increased bid-ask spread as suggested, for example, in \cite{wyart2008relation}. Of course considering the multivariate nature in (\ref{eq:fit_obj}), the resulting pattern depends on the set of variables that we chose at the beginning. Developing other features and selecting the most effective ones in the context of impactful news screening is out of the scope of the current work.    \\

    \section{News screening and learning}
        \label{sec:news_learn}
        \subsection{Methodology}
        \begin{figure}[hbtp]
            \centering
            \includegraphics[width=\textwidth]{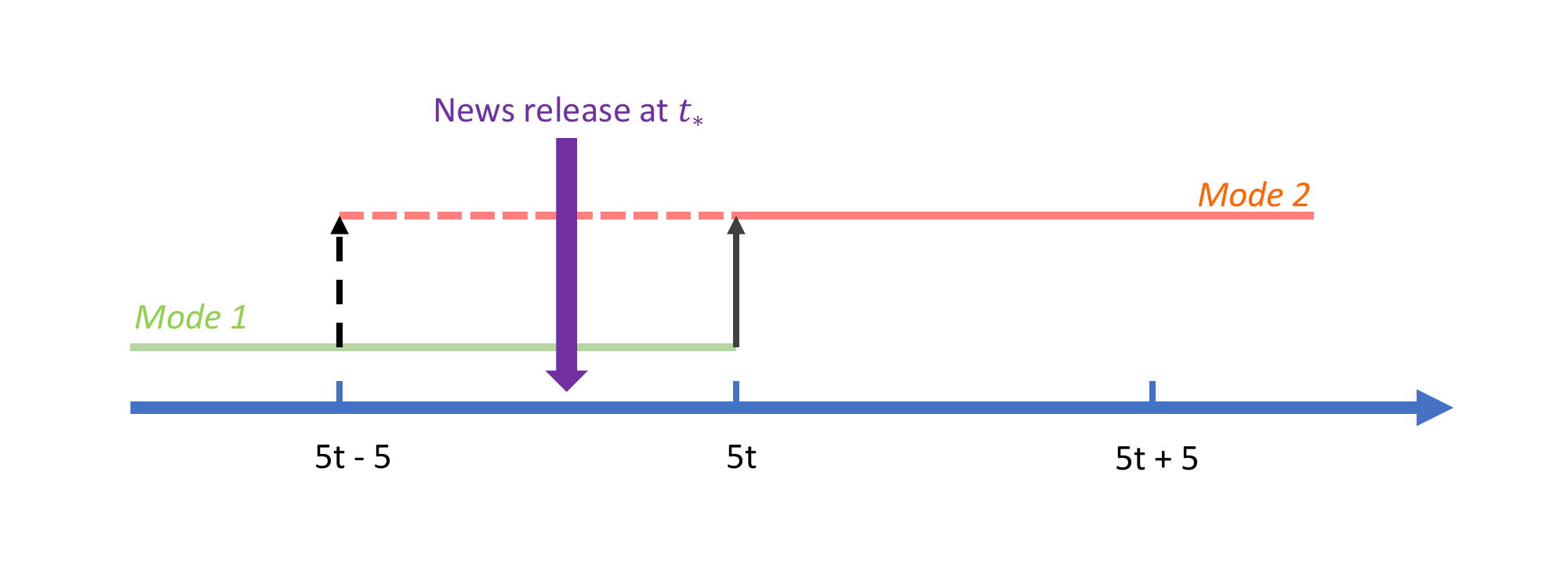}
            \caption{Possible scenarios where a news release is considered impactful.}
            \label{fig:news_scenarios}
        \end{figure}
         It is often suggested in the literature that exogenously driven orders can generate a larger volatility footprint than endogenously mechanical ones, see for example \cite{huang2015simulating, rambaldi2019disentangling}. As \textit{Mode 2} is characterized by higher volatility and increased trading volume, it is reasonable to assume that some exogenous events, news releases in this paper, evoked the mode switches. Considering a piece of news arriving inside the $t$-th 5-minute bin as shown in Figure \ref{fig:news_scenarios}, 
         with $t=1,\dots, 72$, we consider it impactful if one of the following scenarios happens:
         \begin{equation}
             \begin{cases}
                 \text{Jump from \textit{Mode 1} to \textit{Mode 2} at the moment } 5t-5, & \text{for } t \in\{2,\dots,72\} \, , \\
                 \text{Jump from \textit{Mode 1} to \textit{Mode 2} at the moment } 5t, & \text{for } t\in\{1,\dots,71\} \, .
             \end{cases}
            \label{eq:pick_crit}
         \end{equation}\\
        \noindent
        Let the set of original intraday news releases and the impactful ones selected with the above criterion be $\mathcal{D}$ and $\mathcal{D}_{\lambda}$ respectively, where $\lambda$ is the mode switch penalization parameter defined in (\ref{eq:fit_obj}). Given a piece of news $a\in\mathcal{D}$, released at $t^\ast$ and concerning the stock $S$, we denote the $h$-minute return of the stock $S^\ast$ after the publication of $a$ by $r_a^h$, \textit{i.e.} $r_a^h:=\frac{P^{S^\ast}_{t^\ast+h}}{P^{S^\ast}_{t^\ast}} - 1$, where $P^S_t$ represents the price of stock $S$ at time $t$. Since we are interested in the firm-specific price movements disentangled from market-wide activities, we replace $r^h_a$ with its market-detrended version defined by
        $$
            r^h_a := r^h_a - \frac{1}{N}\sum_{S=1}^N( \frac{P^S_{t^\ast+h}}{P^S_{t^\ast}} - 1) \, ,
        $$
        where for sake of simplicity, we follow the classical capital asset pricing model with the \textit{betas} fixed to be one.  
        With $\mathcal{R}^h:=\{r^h_a|a\in\mathcal{D}\}$, let $r^{h,k}$ and $r^{h,100-k}$ be the $k$-th and $(100-k)$-th percentile of $\mathcal{R}^h$ respectively. We define the following sets
        $$
            \mathcal{Z}^{-}_{h,k} := \{a|r^h_a\leq r^{h,k}, a\in\mathcal{D}\} \quad\text{and} \quad \mathcal{Z}^{+}_{h,k}:= \{a|r^h_a\geq r^{h,100-k}, a\in\mathcal{D}\} \, .
        $$
        Thus, $Z^{-}$ and $Z^{+}$ are respectively sets of bearish and bullish news releases according to the amplitude of post-publication stock returns, which is the criterion commonly used in works such as \cite{chen2021stock, ding2014using,ding2015deep,hu2018listening}. The cases with $k \ll 50$ are in the spirit that ``true'' impactful news releases are more likely to be followed with significant price movements. Given $\lambda, h$ and $k$, the sets of ``true'' bearish and bullish news in our approach are defined respectively by
        $$
            \mathcal{N}^{-}_{\lambda, h, k}:=\mathcal{D}_{\lambda} \cap \mathcal{Z}^{-}_{h,k}\quad \text{and}\quad \mathcal{N}^{+}_{\lambda, h, k}:=\mathcal{D}_{\lambda} \cap \mathcal{Z}^{+}_{h,k} \, .
        $$
         Therefore, in addition to extreme post-publication return, the news publications selected by our method is also followed by noticeable changes in market liquidity conditions.  \\
        
        \noindent
        Given a set of labeled samples, such as $\mathcal{E} = \mathcal{Z}^+\cup\mathcal{Z}^-$ or $\mathcal{E} = \mathcal{N}^+\cup\mathcal{N}^-$, we firstly inquire the dependence of news sentiment, positive or negative, on the presence or absence of each individual word through measuring the mutual information between these two random variables. Then we fit a multi-variate Bernoulli NBC as a news sentiment predictor. Some key computational rules are recalled in Appendix 
        \ref{app:nbc}. We refer to for instance \cite{cover1991entropy} and \cite{mccallum1998comparison} for more details of these methods. For a piece of news $a$, the classifier tells us the probabilities of $a$ being positive and negative. Since there is no ground truth for the news sentiment, we evaluate the classification results based on stocks' post-news price movements, under the hypothesis that news sentiment is positively correlated with the direction of stocks' future return. More precisely, for an NBC fitted on the training set $\mathcal{E}$, we define the sentiment score $F\in[-1, 1]$ of news $a$ by
        $$
            F(a)|\mathcal{E} := P^+_{\mathcal{E}}(a) - P^-_{\mathcal{E}}(a) \, ,
        $$ 
        where $P_{\mathcal{E}}^{+}$ and $P^{-}_{\mathcal{E}}$ are the probabilities of $a$ being bullish and bearish respectively given by the model. We expect naturally that a significantly high (low) F is more likely to be followed by a positive (negative) price drift for the associated stock. 

    \subsection{Numerical results}
\begin{figure}[htbp]
            \centering
            \includegraphics[width=.8\linewidth]{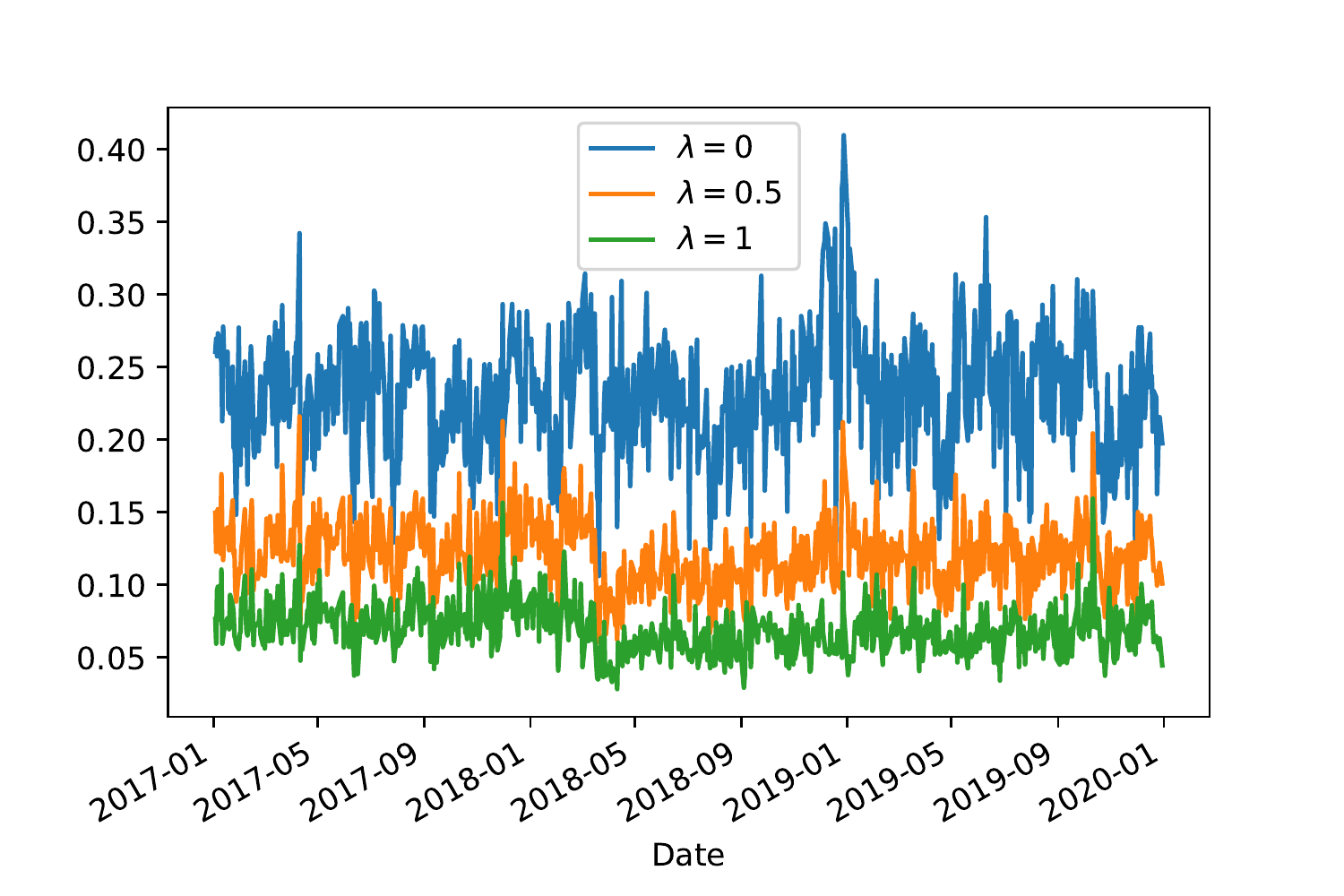}
            \caption{Ratio of news releases with identified nearby jumps from \textit{Mode 1} to \textit{Mode 2}, \textit{i.e.} $\frac{\#\mathcal{D}^{\lambda}}{\#\mathcal{D}}$.}
            \label{fig:news_select}
        \end{figure}

        In this part, we perform news sentiment learning on the screened news dataset.  Figure \ref{fig:news_select} gives the resulting ratio of news releases selected with the criterion (\ref{eq:pick_crit}). Note that with $\lambda=0.5$, only around $10\%$ of total intraday releases are thought to have impacted the market in our approach. 
        
        \subsubsection{Most informative words}
            \begin{figure}[htbp]
                \centering
                \begin{subfigure}{.5\textwidth}
                    \centering
                    \includegraphics[width=\linewidth]{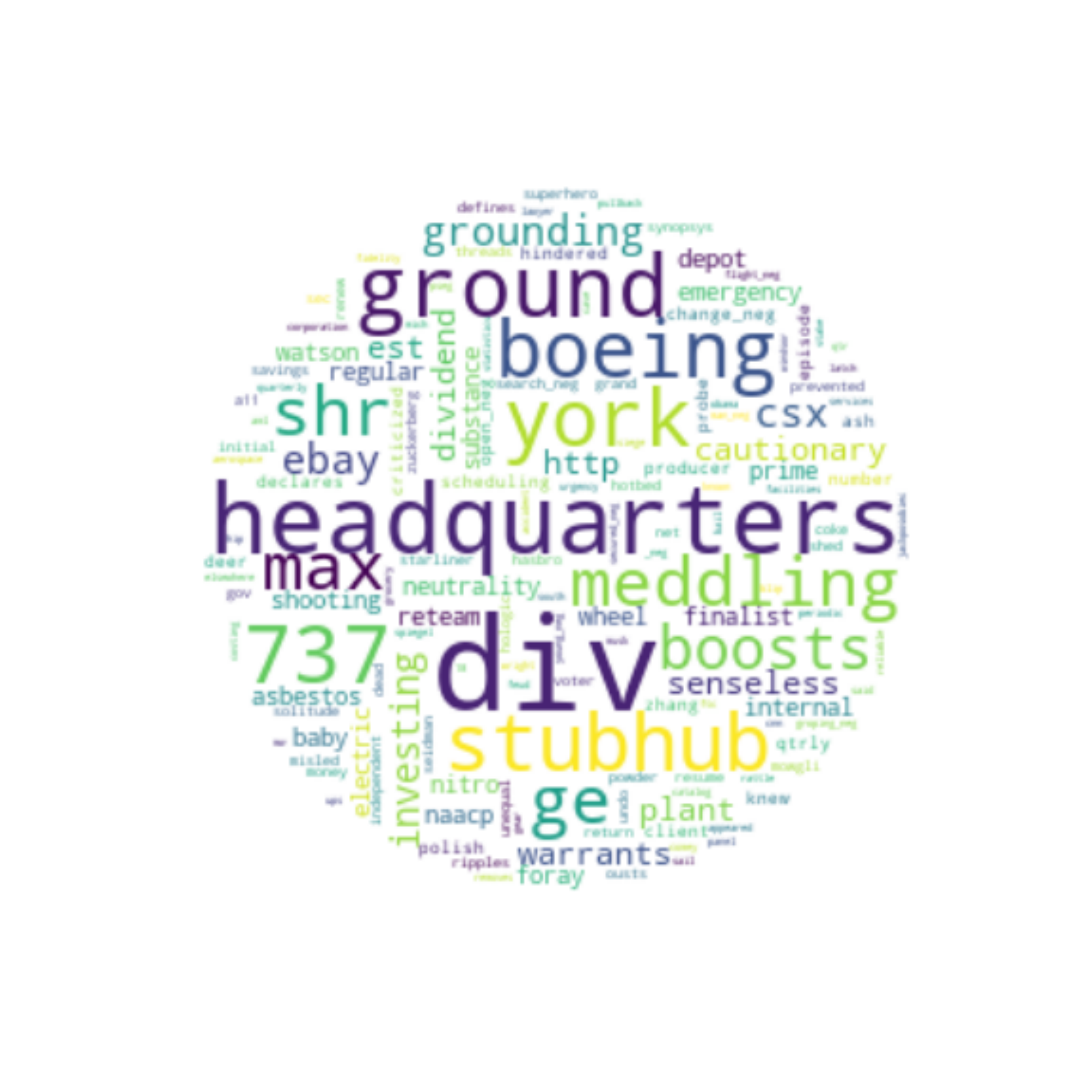}
                \end{subfigure}%
                \begin{subfigure}{.5\textwidth}
                    \centering
                    \includegraphics[width=\linewidth]{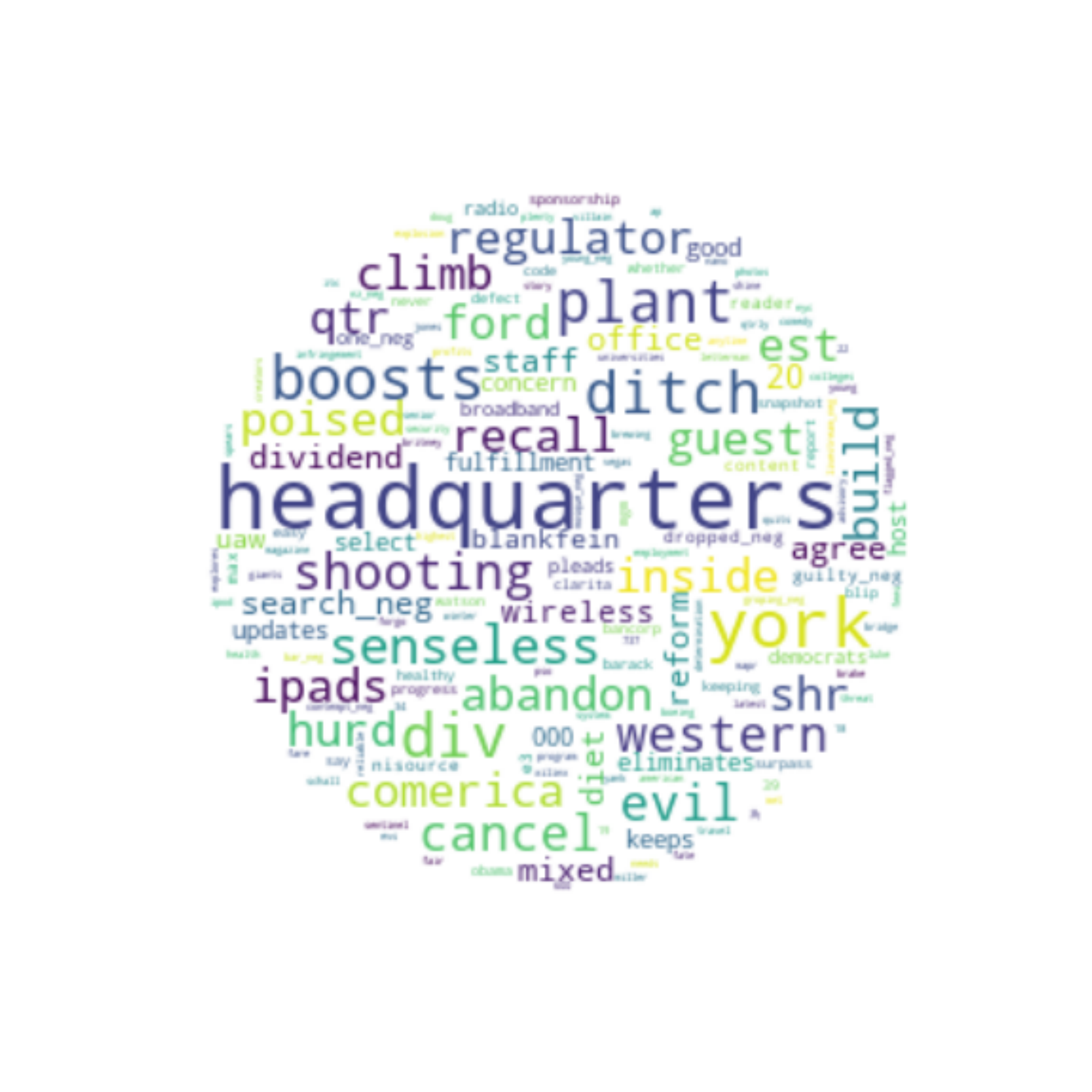}
                \end{subfigure}
                \caption{Words showing the most mutual information with news class variable measured for the samples of $\mathcal{Z}^+_{15,10}\cup\mathcal{Z}^{-}_{15,10}$ (left), and $\mathcal{N}^+_{0.5, 15,10}\cup\mathcal{N}^{-}_{0.5, 15, 10}$ (right). The font size of a word is proportional to the amount of its mutual information.}
                \label{fig:word_cloud}
            \end{figure}

            \noindent
            Figure \ref{fig:word_cloud} shows the words reporting the most mutual information with news sentiment, measured respectively on $\mathcal{Z}^+_{15,10}\cup\mathcal{Z}^{-}_{15,10}$ and $\mathcal{N}^+_{0.5, 15,10}\cup\mathcal{N}^{-}_{0.5, 15, 10}$. Visually, our news selection method boosted with liquidity-driven variables can reduce the weights of certain sentiment-neutral words, \textit{e.g.} \textit{``boeing"}, \textit{``737"}, \textit{``stubhub"}, \textit{``http"}, etc. It also highlights some sentiment-charged words, \textit{e.g.} \textit{``senseless''}, \textit{``abandon''}, \textit{``recall''}, \textit{``poised''}, \textit{``evil''}, etc. Considering the limitation of uni-gram evaluation, it is not surprising that several neutral words still seem to be overvalued with our approach. For example, \textit{``headquarters''} solely is not sentiment-charged, while \textit{``build new headquarters''} is likely to drive some stock price movements.

        \subsubsection{Short-term return prediction}
        \begin{figure}[htbp]
            \centering
            \includegraphics[width=\linewidth]{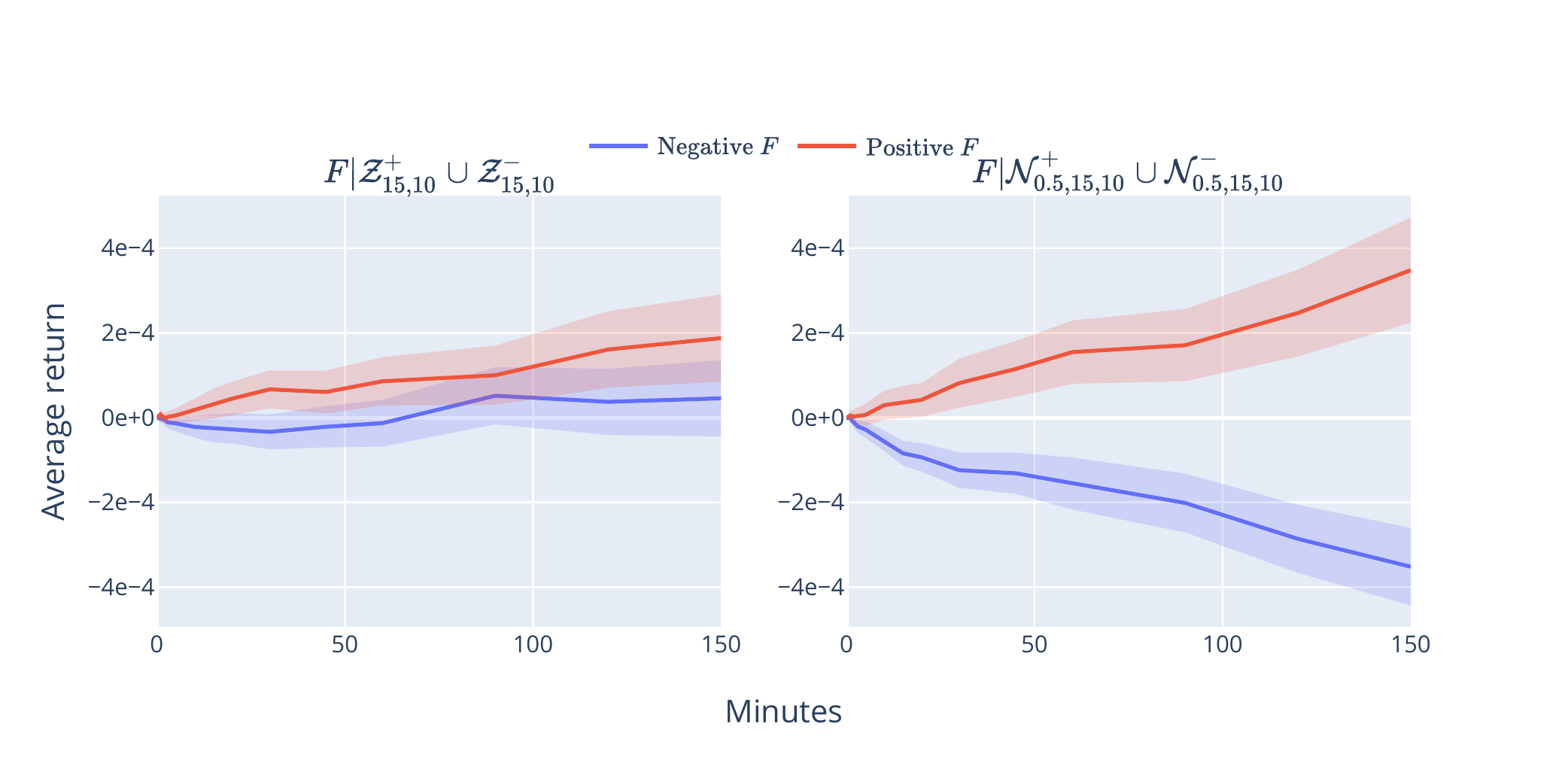}
            \caption{Average post-publication stock return for the news releases with significantly positive and negative sentiment scores assigned by the NBCs fitted on different training sets. The left represents the case with news screening solely based on return information. The right corresponds to our approach. Only the samples whose scores fall in the top/bottom 10$\%$ range are shown for ease of comparison. The shadow parts represent the empirical standard deviation of the average return.}
            \label{fig:news_ret_drift}
        \end{figure}
        
        \noindent
        Now we evaluate the predictive power of the news sentiment scores, given by the fitted NBCs, for future price movement. Note that the following results are based on the news headlines published during the out-of-sample period, \textit{i.e.} 2020-01-01 $\sim$ 2021-12-31. We are more interested in the short-term returns following news publications since they reflect better the immediate impacts of news sentiment compared to price changes over longer horizons. We fit two NBCs on  the training sets $\mathcal{Z}^{+}_{15,10}\cup\mathcal{Z}^{-}_{15,10}$ and $\mathcal{N}^{+}_{0.5,15,10}\cup\mathcal{N}^{-}_{0.5,15,10}$ respectively. They are then applied to the out-of-sample news data to output sentiment estimations. We plot the average post-publication price drifts of the news releases associated with distinct sentiment scores up to 150 minutes in Figure \ref{fig:news_ret_drift}.
        Clearly, after screening the news dataset with our approach, NBC can better predict the short-term stock return. The pieces of news with significantly positive sentiment scores are followed by climbing prices, while the ones with negative scores drive the inverse phenomenon.  \\
        
        \begin{figure}[htbp]
            \centering
            \includegraphics[width=\linewidth]{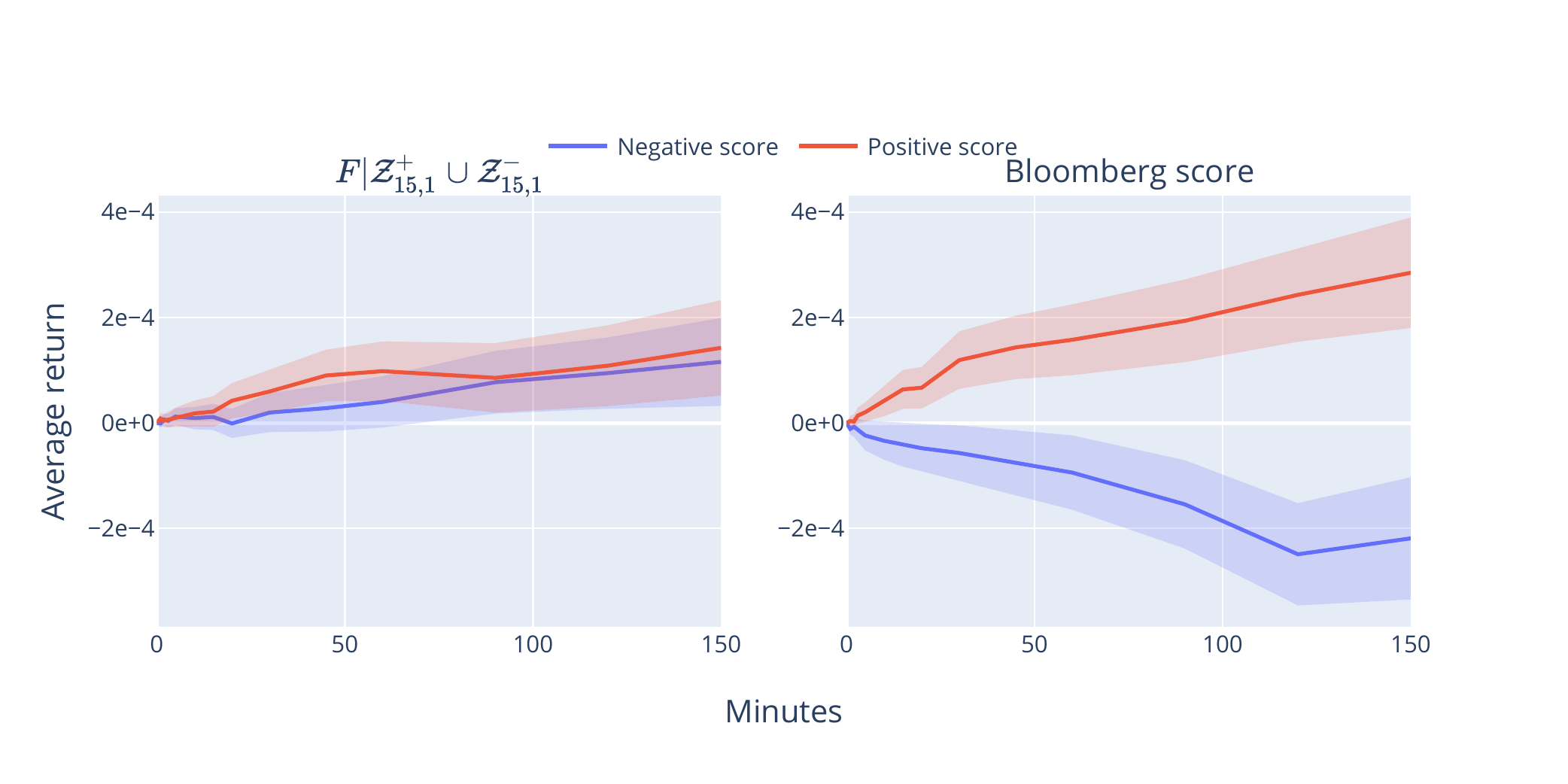}
            \caption{Average post-publication stock return for the news releases with significantly positive and negative sentiment scores assigned by the NBC fitted on the news samples associated with extreme realized return falling in the top/bottom 1$\%$ quantile (left), and the case with the sentiment scores given by Bloomberg (right).}
            \label{fig:news_drift_ret_extreme}
        \end{figure}

        \noindent
        With $\lambda=0.5$, the size of $ \mathcal{N}^{+}_{0.5,15,10}\cup\mathcal{N}^{-}_{0.5,15,10}$ is only about one tenth of $\mathcal{Z}^{+}_{15,10}\cup\mathcal{Z}^{-}_{15,10}$. To verify that our selection method is not equivalent to filtering news releases by the associated post-publicaiton stock returns, we repeat the same test on $\mathcal{Z}^{+}_{15,1}\cup\mathcal{Z}^{-}_{15,1}$. As shown in the left subfigure of Figure \ref{fig:news_drift_ret_extreme}, with similar number of training samples to the case with $\mathcal{N}^{+}_{0.5,15,10}\cup\mathcal{N}^{-}_{0.5,15,10}$, the prediction performance now is largely degraded. Keeping only the news releases with extreme post-publication stock returns can reduce the ratio of neutral samples in the training set to some extent, while it can also result in model underfitting because of data shortage. Our method presents a more efficient way to filter neutral news samples and identify the impactful ones, which helps the model learn more effectively. In Figure \ref{fig:news_drift_ret_extreme}, we show also the results of Bloomberg \textit{composed score} as defined in Section \ref{sec:data}. Interestingly, despite its simplicity, the NBC fitted on the screened dataset performs even slightly better than the scores given by Bloomberg in terms of short-term return prediction. Moreover, we show in Appendix \ref{app:robust} that the performance of our approach is not very sensitive to the value of $\lambda$, $h$ and $k$.

    \section{Conclusion}
        \label{sec:conc}
        In this work, we introduce a systematic method for identifying ``true'' impactful news releases that are conceived to contain unexpected information for the financial market.
        The identification method consists in associating significant changes in liquidity conditions during market open hours with nearby news publications. Four variables are used to monitor the intraday dynamics of liquidity mode, including volatility, turnover, bid-ask spread, and book size. Through numerical tests on S\&P500 components, the two predetermined liquidity states are distinct from each other in terms of volatility and turnover level. We pick out the news releases with identifiable mode switch from that with lower volatility and less trading volume to the other one, and label them by the sign of post-news realized returns of the associated stocks. Experiments on news sentiment learning with NBC show that the proposed news screening method leads to more effective feature capturing and thus better model predictive performance. \\

        \noindent
        The study can be extended in several directions. First, our news screening approach can be taken as a preprocessing procedure, and can be applied together with various news sentiment learning methods. Second, we focus on the Bloomberg News data in this study, while the same tests can be easily conducted on any other news dataset, or more generally on other types of exogenous events/signals. Last, it would be interesting to build some agent-based models to understand further the link between exogenous inputs and resulting dynamics of volatility/turnover.

    \clearpage
    \begin{appendices}
        \section{Jump model fitting}
        \label{app:jump_algo}

        For each day, we run the following algorithm:
        \begin{algorithm}
            \caption{Fitting algorithm for the problem (\ref{eq:fit_obj})}
            \label{algo:fit}
            \KwInput{$N$ observations sequences $(x^{S}_t)_{t=1,\dots,T}$, with $S=1,\dots, N$, jump penalty $\lambda$ and convergence tolerance $\epsilon$.}
            \begin{enumerate}
                \item Generate $N$ initial mode sequences $\{(m^{S}_t)_{t=1,\dots,T}\}_{S=1,\dots,N}$ through applying K-means on the pooled set of observation sequences. Initial loss $J^0=+\infty$. 
                \item Iterate for $l=1,\dots$ until $|J^l - J^{l-1}|\leq \epsilon$: 
                    \begin{enumerate}
                        \item Model parameter fit: for $k=1,\dots,K$, the optimal $\theta_k$ is given by the mean of all the samples assigned with mode $k$, \textit{i.e.} 
                        $$\theta_k=\frac{\sum_{S=1}^N\sum_{t=1}^{T}x^{S}_t\mathbbm{1}_{m^S_t=k}}{\sum_{S=1}^N\sum_{t=1}^{T}\mathbbm{1}_{m^S_t=k}}, \quad i=1,\cdots, 4 $$.
                        \item For each $S=1,\dots, N$, solve the optimal mode sequence with respect to $\theta_1, \theta_2$:
                            \begin{enumerate}
                                \item Compute a matrix $F^S\in \mathbb{R}^{T\times K}$, which is defined by:
                                    \begin{align*}
                                        F^S(T, k) &= \|x^S_T - \theta_k\|_2^2\, , \\
                                        F^S(t, k) &= \|x^S_t - \theta_k\|_2^2 + \min_j\{F^S(t+1, j) + \lambda \mathbbm{1}_{k\neq j}\}\, ,  
                                    \end{align*}
                                    where $k=1,\dots,K$. 
                                \item Reconstruct the optimal mode sequence with
                                    \begin{align*}
                                        m^S_1 &= \arg\min_k F^S(1, k) \, , \\
                                        m^S_t &= \arg\min_k \{F^S(t, k) + \lambda\mathbbm{1}_{m^S_{t-1}\neq k}\} \, , \quad t=2,\dots,T \,.
                                    \end{align*}
                            \end{enumerate} 
                        \item Update the loss by $J^l=\sum_{S=1}^NF^S(1, m^S_1)$.
                    \end{enumerate}
            \end{enumerate}
            \KwOutput{Model parameters $\theta_1,\dots, \theta_K$, $N$ mode sequences $\{(m^{S}_t)_{t=1,\dots,T}\}_{S=1,\dots,N}$.}    
        \end{algorithm}
        \vskip 0.15in
        \noindent
        At each iteration the loss $J$ is non-increasing, so Algorithm \ref{algo:fit} can always terminate in a finite number of steps. However, similar to other classical clustering 
        methods such as K-means or Gaussian mixture methods, the final solution depends on the initialization and may not be the global optimal one. In practice, one can run the above algorithm multiple times 
        with different initial model sequences and keep the one with the smallest loss.    

        \section{Mutual information and naive Bayes classifier}
        \label{app:nbc}
        \textbf{Mutual information} \\

        \noindent
        Let $C\in\mathcal{C}$ and $W\in\{0, 1\}$ be two random variables denoting respectively the category of a news release, \textit{i.e.} positive or negative in this paper, and the presence or absence of word $w$ in the news text. $W=0$ represents the absence of the word, and $W=1$ represents the presence of $w$. The mutual information between $C$ and $W$ is given by
        $$
            I(C;W) = \sum_{c\in\mathcal{C}}\sum_{f_w\in\{0, 1\}} P_{(C,W)}(c, f_w)\log \Big(\frac{P_{(C,W)}(c,f_w)}{P_C(c)P_W(f_w)}\Big) \, ,
        $$
        where $P_C$ and $P_W$ are the marginal probability mass functions of $C$ and $W$ respectively, and $P_{(C,W)}$ is their joint probability mass function. In this work, for a given labeled set of news headlines, $\mathcal{Z}^{-}\cup\mathcal{Z}^{+}$ or $\mathcal{N}^{-}\cup\mathcal{N}^{+}$, we estimate the above probability quantities by their classical empirical observations over all samples.  \\  

        \noindent
        \textbf{Multi-variate Bernoulli naive Bayes classifier} \\
        \noindent
        Let $c$ be the news class variable, $f_{w_1}, \dots, f_{w_n}$ indicate the presence or absence of word $w_1, \dots, w_n$ in the news headlines, we are interested in the classification result given by
        $$
            \hat{c} = \arg\max_c P(c|f_{w_1},\dots, f_{w_n}) \, .
        $$
        Following Bayes' theorem, we have
        $$
            P(c|f_{w_1},\dots, f_{w_n}) = \frac{P(c)P(f_{w_1},\dots,f_{w_n}|c)}{P(f_{w_1},\dots,f_{w_n})} \, .
        $$
        The ``naive'' conditional independence assumption, we have
        $$
            P(f_{w_1},\dots,f_{w_n}|c) = \prod_{i=1}^n P(f_{w_i}|c) \, ,
        $$
        where $P(f_{w_i}|c)$ can be easily estimated given the Bernoulli assumption.

        \section{Robusteness tests}
        \label{app:robust}
        Our approach involves mainly three parameters, \textit{i.e.} $\lambda$, $h$ and $k$. In the following we show their effects by varying their values.  \\

        \noindent
        \textbf{- Effect of $\lambda$} \\

\begin{figure}[htbp]
            \centering
            \includegraphics[width=\textwidth]{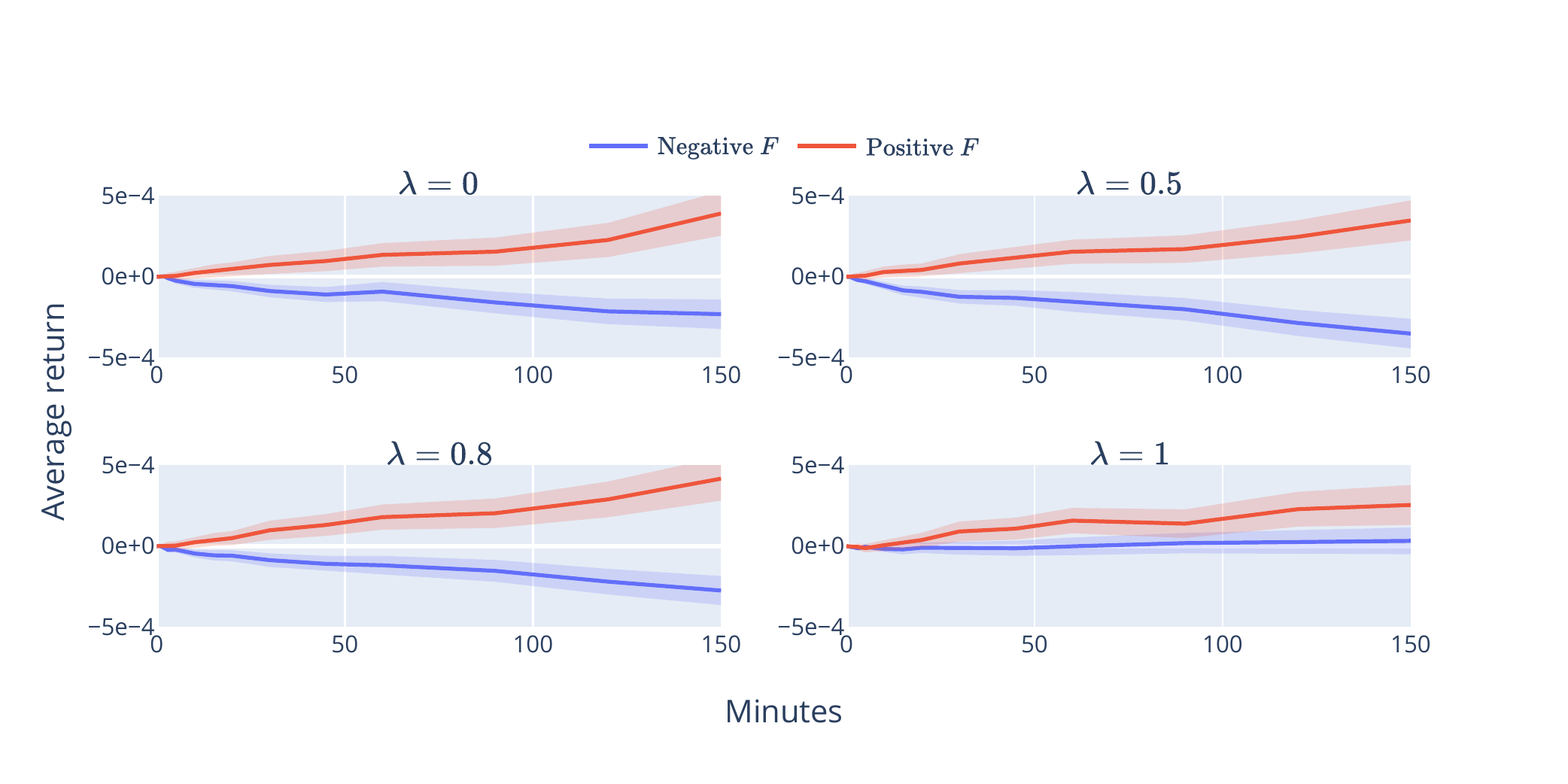}
            \caption{Average post-news stock return corresponding to the cases with different jump penalization parameter $\lambda$. We set $h=15$, $k=10$.}
            \label{fig:effect_lambda}
        \end{figure}

        % \vskip 0.05in
        \noindent
        As shown in Figure \ref{fig:effect_lambda}, the results are relatively robust with intermediate $\lambda$. When $\lambda=1$, the amount of news samples taken as impactful becomes too limited to accurate model learning. \\

        \noindent
        \textbf{- Effect of $h$} \\
        % \vskip 0.05in

\begin{figure}[htbp]
            \centering
            \includegraphics[width=\textwidth]{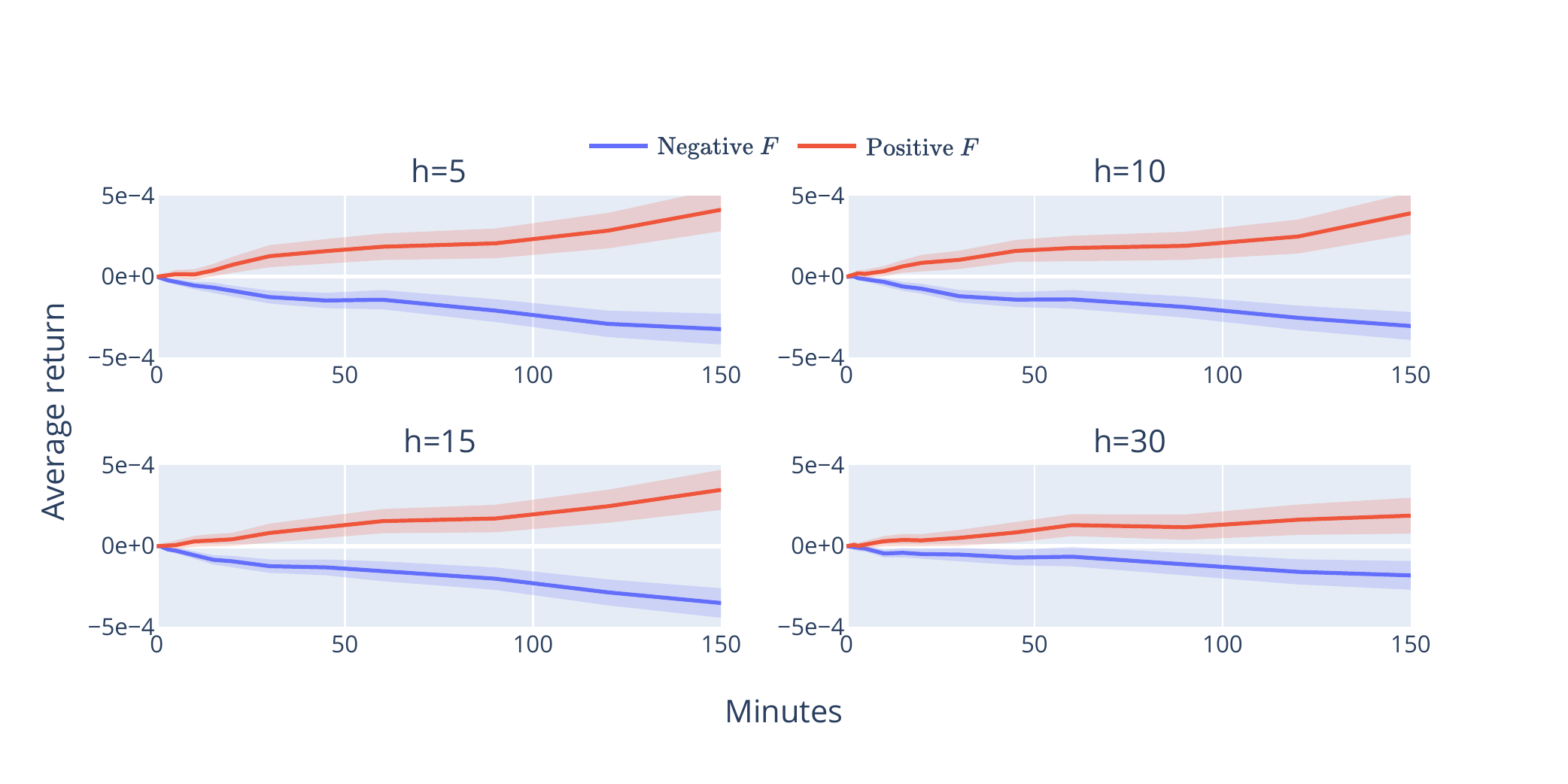}
            \caption{Average post-news stock return corresponding to the cases with different $h$. We set $\lambda=0.5$, $k=10$.}
            \label{fig:effect_horizon}
        \end{figure}

        \noindent
        Figure \ref{fig:effect_horizon} gives the results when choosing the post-news realized return over different horizons for news sign labeling. We do not remark significant variation of performance for $h\in[5, 10, 15]$. Slight degradation is observed for $h=30$, which is understandable given the increased volatility of realized return. \\

        \noindent
        \textbf{- Effect of $k$} \\

\begin{figure}[htbp]
            \centering
            \includegraphics[width=\textwidth]{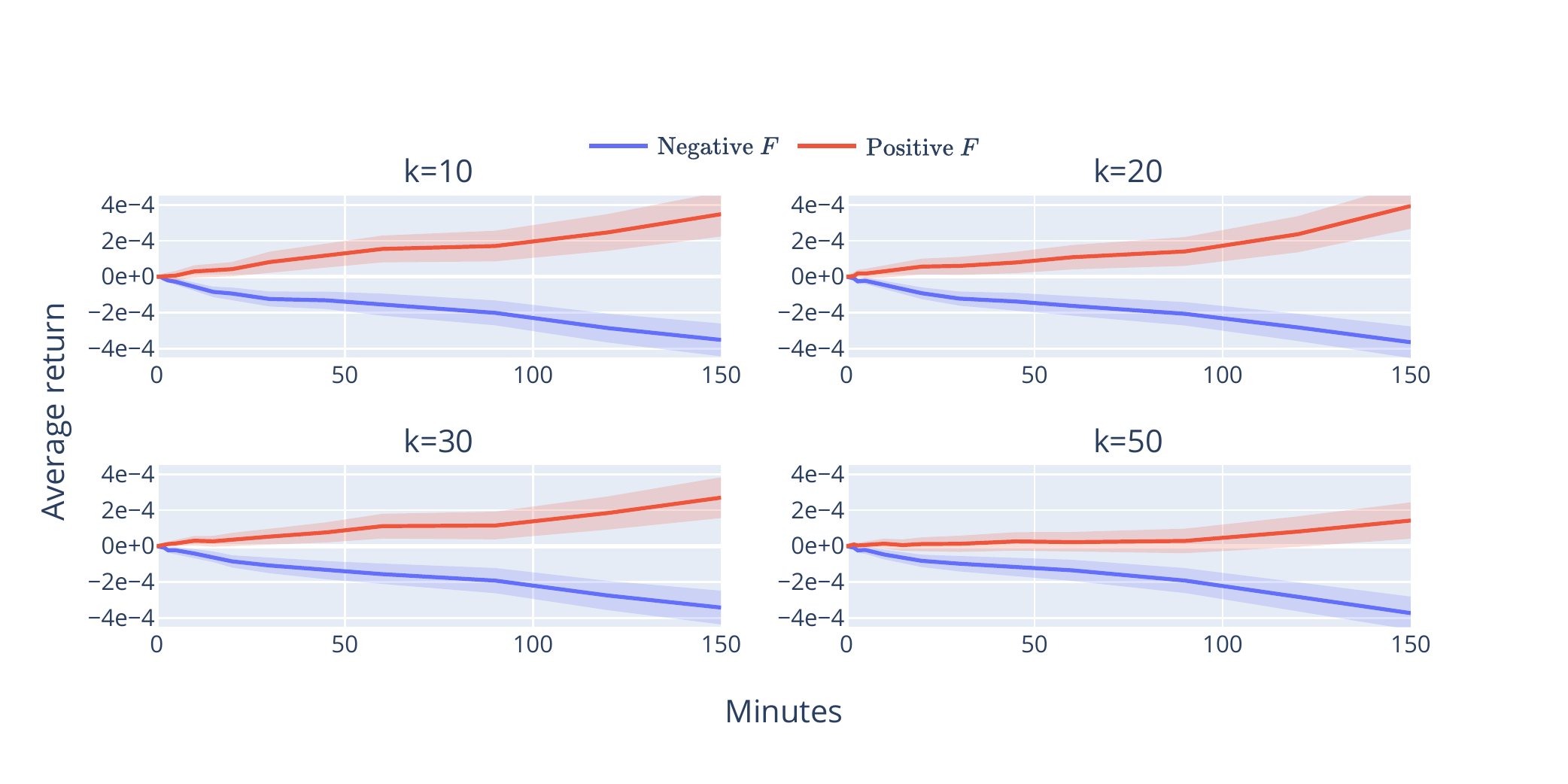}
            \caption{Average post-news stock return corresponding to the cases with different $k$. We set $\lambda=0.5$, $h=15$.}
            \label{fig:effect_pct}    
        \end{figure}

        \noindent
        When using only realized return information for news labeling, we are inclined to relatively small $k$ to pick out only the news releases associated with significant post-publication price drifts. With our method, as shown in Figure \ref{fig:effect_pct}, the impact of $k$ on the final results is very limited. Even with $k=50$, which means that the magnitude of realized return is not considered, there is no significant deterioration for the predictive performance of the resulting model. \\

    \end{appendices}
    \clearpage
    \bibliography{ref}

\begin{thebibliography}{}

\bibitem [\protect \citeauthoryear {%
Bemporad%
, Breschi%
, Piga%
\BCBL {}\ \BBA {} Boyd%
}{%
Bemporad%
\ \protect \BOthers {.}}{%
{\protect \APACyear {2018}}%
}]{%
bemporad2018fitting}
\APACinsertmetastar {%
bemporad2018fitting}%
\begin{APACrefauthors}%
Bemporad, A.%
, Breschi, V.%
, Piga, D.%
\BCBL {}\ \BBA {} Boyd, S\BPBI P.%
\end{APACrefauthors}%
\unskip\
\newblock
\APACrefYearMonthDay{2018}{}{}.
\newblock
{\BBOQ}\APACrefatitle {Fitting jump models} {Fitting jump models}.{\BBCQ}
\newblock
\APACjournalVolNumPages{Automatica}{96}{}{11--21}.
\PrintBackRefs{\CurrentBib}

\bibitem [\protect \citeauthoryear {%
Bi{\'n}kowski%
\ \BBA {} Lehalle%
}{%
Bi{\'n}kowski%
\ \BBA {} Lehalle%
}{%
{\protect \APACyear {2022}}%
}]{%
binkowski2022endogenous}
\APACinsertmetastar {%
binkowski2022endogenous}%
\begin{APACrefauthors}%
Bi{\'n}kowski, M.%
\BCBT {}\ \BBA {} Lehalle, C\BHBI A.%
\end{APACrefauthors}%
\unskip\
\newblock
\APACrefYearMonthDay{2022}{}{}.
\newblock
{\BBOQ}\APACrefatitle {Endogenous Dynamics of Intraday Liquidity} {Endogenous
  dynamics of intraday liquidity}.{\BBCQ}
\newblock
\APACjournalVolNumPages{The Journal of Portfolio Management}{48}{6}{145--169}.
\PrintBackRefs{\CurrentBib}

\bibitem [\protect \citeauthoryear {%
Chan%
}{%
Chan%
}{%
{\protect \APACyear {2003}}%
}]{%
chan2003stock}
\APACinsertmetastar {%
chan2003stock}%
\begin{APACrefauthors}%
Chan, W\BPBI S.%
\end{APACrefauthors}%
\unskip\
\newblock
\APACrefYearMonthDay{2003}{}{}.
\newblock
{\BBOQ}\APACrefatitle {Stock price reaction to news and no-news: drift and
  reversal after headlines} {Stock price reaction to news and no-news: drift
  and reversal after headlines}.{\BBCQ}
\newblock
\APACjournalVolNumPages{Journal of Financial Economics}{70}{2}{223--260}.
\PrintBackRefs{\CurrentBib}

\bibitem [\protect \citeauthoryear {%
Chen%
}{%
Chen%
}{%
{\protect \APACyear {2021}}%
}]{%
chen2021stock}
\APACinsertmetastar {%
chen2021stock}%
\begin{APACrefauthors}%
Chen, Q.%
\end{APACrefauthors}%
\unskip\
\newblock
\APACrefYearMonthDay{2021}{}{}.
\newblock
{\BBOQ}\APACrefatitle {Stock movement prediction with financial news using
  contextualized embedding from bert} {Stock movement prediction with financial
  news using contextualized embedding from bert}.{\BBCQ}
\newblock
\APACjournalVolNumPages{arXiv preprint arXiv:2107.08721}{}{}{}.
\PrintBackRefs{\CurrentBib}

\bibitem [\protect \citeauthoryear {%
Cover%
, Thomas%
\BCBL {}\ \protect \BOthers {.}}{%
Cover%
\ \protect \BOthers {.}}{%
{\protect \APACyear {1991}}%
}]{%
cover1991entropy}
\APACinsertmetastar {%
cover1991entropy}%
\begin{APACrefauthors}%
Cover, T\BPBI M.%
, Thomas, J\BPBI A.%
\BCBL {}\ \BOthersPeriod {.}\end{APACrefauthors}%
\unskip\
\newblock
\APACrefYearMonthDay{1991}{}{}.
\newblock
{\BBOQ}\APACrefatitle {Entropy, relative entropy and mutual information}
  {Entropy, relative entropy and mutual information}.{\BBCQ}
\newblock
\APACjournalVolNumPages{Elements of information theory}{2}{1}{12--13}.
\PrintBackRefs{\CurrentBib}

\bibitem [\protect \citeauthoryear {%
Ding%
, Zhang%
, Liu%
\BCBL {}\ \BBA {} Duan%
}{%
Ding%
\ \protect \BOthers {.}}{%
{\protect \APACyear {2014}}%
}]{%
ding2014using}
\APACinsertmetastar {%
ding2014using}%
\begin{APACrefauthors}%
Ding, X.%
, Zhang, Y.%
, Liu, T.%
\BCBL {}\ \BBA {} Duan, J.%
\end{APACrefauthors}%
\unskip\
\newblock
\APACrefYearMonthDay{2014}{}{}.
\newblock
{\BBOQ}\APACrefatitle {Using structured events to predict stock price movement:
  An empirical investigation} {Using structured events to predict stock price
  movement: An empirical investigation}.{\BBCQ}
\newblock
\BIn{} \APACrefbtitle {Proceedings of the 2014 conference on empirical methods
  in natural language processing (EMNLP)} {Proceedings of the 2014 conference
  on empirical methods in natural language processing (emnlp)}\ (\BPGS\
  1415--1425).
\PrintBackRefs{\CurrentBib}

\bibitem [\protect \citeauthoryear {%
Ding%
, Zhang%
, Liu%
\BCBL {}\ \BBA {} Duan%
}{%
Ding%
\ \protect \BOthers {.}}{%
{\protect \APACyear {2015}}%
}]{%
ding2015deep}
\APACinsertmetastar {%
ding2015deep}%
\begin{APACrefauthors}%
Ding, X.%
, Zhang, Y.%
, Liu, T.%
\BCBL {}\ \BBA {} Duan, J.%
\end{APACrefauthors}%
\unskip\
\newblock
\APACrefYearMonthDay{2015}{}{}.
\newblock
{\BBOQ}\APACrefatitle {Deep learning for event-driven stock prediction} {Deep
  learning for event-driven stock prediction}.{\BBCQ}
\newblock
\BIn{} \APACrefbtitle {Twenty-fourth international joint conference on
  artificial intelligence.} {Twenty-fourth international joint conference on
  artificial intelligence.}
\PrintBackRefs{\CurrentBib}

\bibitem [\protect \citeauthoryear {%
Gro{\ss}-Klu{\ss}mann%
\ \BBA {} Hautsch%
}{%
Gro{\ss}-Klu{\ss}mann%
\ \BBA {} Hautsch%
}{%
{\protect \APACyear {2011}}%
}]{%
gross2011machines}
\APACinsertmetastar {%
gross2011machines}%
\begin{APACrefauthors}%
Gro{\ss}-Klu{\ss}mann, A.%
\BCBT {}\ \BBA {} Hautsch, N.%
\end{APACrefauthors}%
\unskip\
\newblock
\APACrefYearMonthDay{2011}{}{}.
\newblock
{\BBOQ}\APACrefatitle {When machines read the news: Using automated text
  analytics to quantify high frequency news-implied market reactions} {When
  machines read the news: Using automated text analytics to quantify high
  frequency news-implied market reactions}.{\BBCQ}
\newblock
\APACjournalVolNumPages{Journal of Empirical Finance}{18}{2}{321--340}.
\PrintBackRefs{\CurrentBib}

\bibitem [\protect \citeauthoryear {%
Hu%
, Liu%
, Bian%
, Liu%
\BCBL {}\ \BBA {} Liu%
}{%
Hu%
\ \protect \BOthers {.}}{%
{\protect \APACyear {2018}}%
}]{%
hu2018listening}
\APACinsertmetastar {%
hu2018listening}%
\begin{APACrefauthors}%
Hu, Z.%
, Liu, W.%
, Bian, J.%
, Liu, X.%
\BCBL {}\ \BBA {} Liu, T\BHBI Y.%
\end{APACrefauthors}%
\unskip\
\newblock
\APACrefYearMonthDay{2018}{}{}.
\newblock
{\BBOQ}\APACrefatitle {Listening to chaotic whispers: A deep learning framework
  for news-oriented stock trend prediction} {Listening to chaotic whispers: A
  deep learning framework for news-oriented stock trend prediction}.{\BBCQ}
\newblock
\BIn{} \APACrefbtitle {Proceedings of the eleventh ACM international conference
  on web search and data mining} {Proceedings of the eleventh acm international
  conference on web search and data mining}\ (\BPGS\ 261--269).
\PrintBackRefs{\CurrentBib}

\bibitem [\protect \citeauthoryear {%
Huang%
, Lehalle%
\BCBL {}\ \BBA {} Rosenbaum%
}{%
Huang%
\ \protect \BOthers {.}}{%
{\protect \APACyear {2015}}%
}]{%
huang2015simulating}
\APACinsertmetastar {%
huang2015simulating}%
\begin{APACrefauthors}%
Huang, W.%
, Lehalle, C\BHBI A.%
\BCBL {}\ \BBA {} Rosenbaum, M.%
\end{APACrefauthors}%
\unskip\
\newblock
\APACrefYearMonthDay{2015}{}{}.
\newblock
{\BBOQ}\APACrefatitle {Simulating and analyzing order book data: The
  queue-reactive model} {Simulating and analyzing order book data: The
  queue-reactive model}.{\BBCQ}
\newblock
\APACjournalVolNumPages{Journal of the American Statistical
  Association}{110}{509}{107--122}.
\PrintBackRefs{\CurrentBib}

\bibitem [\protect \citeauthoryear {%
Jiang%
, Li%
\BCBL {}\ \BBA {} Wang%
}{%
Jiang%
\ \protect \BOthers {.}}{%
{\protect \APACyear {2021}}%
}]{%
jiang2021pervasive}
\APACinsertmetastar {%
jiang2021pervasive}%
\begin{APACrefauthors}%
Jiang, H.%
, Li, S\BPBI Z.%
\BCBL {}\ \BBA {} Wang, H.%
\end{APACrefauthors}%
\unskip\
\newblock
\APACrefYearMonthDay{2021}{}{}.
\newblock
{\BBOQ}\APACrefatitle {Pervasive underreaction: Evidence from high-frequency
  data} {Pervasive underreaction: Evidence from high-frequency data}.{\BBCQ}
\newblock
\APACjournalVolNumPages{Journal of Financial Economics}{141}{2}{573--599}.
\PrintBackRefs{\CurrentBib}

\bibitem [\protect \citeauthoryear {%
Joulin%
, Lefevre%
, Grunberg%
\BCBL {}\ \BBA {} Bouchaud%
}{%
Joulin%
\ \protect \BOthers {.}}{%
{\protect \APACyear {2008}}%
}]{%
joulin2008stock}
\APACinsertmetastar {%
joulin2008stock}%
\begin{APACrefauthors}%
Joulin, A.%
, Lefevre, A.%
, Grunberg, D.%
\BCBL {}\ \BBA {} Bouchaud, J\BHBI P.%
\end{APACrefauthors}%
\unskip\
\newblock
\APACrefYearMonthDay{2008}{}{}.
\newblock
{\BBOQ}\APACrefatitle {Stock price jumps: news and volume play a minor role}
  {Stock price jumps: news and volume play a minor role}.{\BBCQ}
\newblock
\APACjournalVolNumPages{Wilmott Magazine}{}{46}{}.
\PrintBackRefs{\CurrentBib}

\bibitem [\protect \citeauthoryear {%
Ke%
, Kelly%
\BCBL {}\ \BBA {} Xiu%
}{%
Ke%
\ \protect \BOthers {.}}{%
{\protect \APACyear {2019}}%
}]{%
ke2019predicting}
\APACinsertmetastar {%
ke2019predicting}%
\begin{APACrefauthors}%
Ke, Z\BPBI T.%
, Kelly, B\BPBI T.%
\BCBL {}\ \BBA {} Xiu, D.%
\end{APACrefauthors}%
\unskip\
\newblock
\APACrefYearMonthDay{2019}{}{}.
\newblock
\APACrefbtitle {Predicting returns with text data} {Predicting returns with
  text data}\ \APACbVolEdTR{}{\BTR{}}.
\newblock
\APACaddressInstitution{}{National Bureau of Economic Research}.
\PrintBackRefs{\CurrentBib}

\bibitem [\protect \citeauthoryear {%
Kenton%
\ \BBA {} Toutanova%
}{%
Kenton%
\ \BBA {} Toutanova%
}{%
{\protect \APACyear {2019}}%
}]{%
kenton2019bert}
\APACinsertmetastar {%
kenton2019bert}%
\begin{APACrefauthors}%
Kenton, J\BPBI D\BPBI M\BHBI W\BPBI C.%
\BCBT {}\ \BBA {} Toutanova, L\BPBI K.%
\end{APACrefauthors}%
\unskip\
\newblock
\APACrefYearMonthDay{2019}{}{}.
\newblock
{\BBOQ}\APACrefatitle {Bert: Pre-training of deep bidirectional transformers
  for language understanding} {Bert: Pre-training of deep bidirectional
  transformers for language understanding}.{\BBCQ}
\newblock
\BIn{} \APACrefbtitle {Proceedings of naacL-HLT} {Proceedings of naacl-hlt}\
  (\BPGS\ 4171--4186).
\PrintBackRefs{\CurrentBib}

\bibitem [\protect \citeauthoryear {%
Lehalle%
\ \BBA {} Laruelle%
}{%
Lehalle%
\ \BBA {} Laruelle%
}{%
{\protect \APACyear {2018}}%
}]{%
lehalle2018market}
\APACinsertmetastar {%
lehalle2018market}%
\begin{APACrefauthors}%
Lehalle, C\BHBI A.%
\BCBT {}\ \BBA {} Laruelle, S.%
\end{APACrefauthors}%
\unskip\
\newblock
\APACrefYear{2018}.
\newblock
\APACrefbtitle {Market microstructure in practice} {Market microstructure in
  practice}.
\newblock
\APACaddressPublisher{}{World Scientific}.
\PrintBackRefs{\CurrentBib}

\bibitem [\protect \citeauthoryear {%
Loughran%
\ \BBA {} McDonald%
}{%
Loughran%
\ \BBA {} McDonald%
}{%
{\protect \APACyear {2011}}%
}]{%
loughran2011liability}
\APACinsertmetastar {%
loughran2011liability}%
\begin{APACrefauthors}%
Loughran, T.%
\BCBT {}\ \BBA {} McDonald, B.%
\end{APACrefauthors}%
\unskip\
\newblock
\APACrefYearMonthDay{2011}{}{}.
\newblock
{\BBOQ}\APACrefatitle {When is a liability not a liability? Textual analysis,
  dictionaries, and 10-Ks} {When is a liability not a liability? textual
  analysis, dictionaries, and 10-ks}.{\BBCQ}
\newblock
\APACjournalVolNumPages{The Journal of finance}{66}{1}{35--65}.
\PrintBackRefs{\CurrentBib}

\bibitem [\protect \citeauthoryear {%
Marcaccioli%
, Bouchaud%
\BCBL {}\ \BBA {} Benzaquen%
}{%
Marcaccioli%
\ \protect \BOthers {.}}{%
{\protect \APACyear {2022}}%
}]{%
marcaccioli2022exogenous}
\APACinsertmetastar {%
marcaccioli2022exogenous}%
\begin{APACrefauthors}%
Marcaccioli, R.%
, Bouchaud, J\BHBI P.%
\BCBL {}\ \BBA {} Benzaquen, M.%
\end{APACrefauthors}%
\unskip\
\newblock
\APACrefYearMonthDay{2022}{}{}.
\newblock
{\BBOQ}\APACrefatitle {Exogenous and endogenous price jumps belong to different
  dynamical classes} {Exogenous and endogenous price jumps belong to different
  dynamical classes}.{\BBCQ}
\newblock
\APACjournalVolNumPages{Journal of Statistical Mechanics: Theory and
  Experiment}{2022}{2}{023403}.
\PrintBackRefs{\CurrentBib}

\bibitem [\protect \citeauthoryear {%
McCallum%
, Nigam%
\BCBL {}\ \protect \BOthers {.}}{%
McCallum%
\ \protect \BOthers {.}}{%
{\protect \APACyear {1998}}%
}]{%
mccallum1998comparison}
\APACinsertmetastar {%
mccallum1998comparison}%
\begin{APACrefauthors}%
McCallum, A.%
, Nigam, K.%
\BCBL {}\ \BOthersPeriod {.}\end{APACrefauthors}%
\unskip\
\newblock
\APACrefYearMonthDay{1998}{}{}.
\newblock
{\BBOQ}\APACrefatitle {A comparison of event models for naive bayes text
  classification} {A comparison of event models for naive bayes text
  classification}.{\BBCQ}
\newblock
\BIn{} \APACrefbtitle {AAAI-98 workshop on learning for text categorization}
  {Aaai-98 workshop on learning for text categorization}\ (\BVOL~752, \BPGS\
  41--48).
\PrintBackRefs{\CurrentBib}

\bibitem [\protect \citeauthoryear {%
Nystrup%
, Kolm%
\BCBL {}\ \BBA {} Lindstr{\"o}m%
}{%
Nystrup%
\ \protect \BOthers {.}}{%
{\protect \APACyear {2021}}%
}]{%
nystrup2021feature}
\APACinsertmetastar {%
nystrup2021feature}%
\begin{APACrefauthors}%
Nystrup, P.%
, Kolm, P\BPBI N.%
\BCBL {}\ \BBA {} Lindstr{\"o}m, E.%
\end{APACrefauthors}%
\unskip\
\newblock
\APACrefYearMonthDay{2021}{}{}.
\newblock
{\BBOQ}\APACrefatitle {Feature selection in jump models} {Feature selection in
  jump models}.{\BBCQ}
\newblock
\APACjournalVolNumPages{Expert Systems with Applications}{184}{}{115558}.
\PrintBackRefs{\CurrentBib}

\bibitem [\protect \citeauthoryear {%
Oliveira%
, Cortez%
\BCBL {}\ \BBA {} Areal%
}{%
Oliveira%
\ \protect \BOthers {.}}{%
{\protect \APACyear {2016}}%
}]{%
oliveira2016stock}
\APACinsertmetastar {%
oliveira2016stock}%
\begin{APACrefauthors}%
Oliveira, N.%
, Cortez, P.%
\BCBL {}\ \BBA {} Areal, N.%
\end{APACrefauthors}%
\unskip\
\newblock
\APACrefYearMonthDay{2016}{}{}.
\newblock
{\BBOQ}\APACrefatitle {Stock market sentiment lexicon acquisition using
  microblogging data and statistical measures} {Stock market sentiment lexicon
  acquisition using microblogging data and statistical measures}.{\BBCQ}
\newblock
\APACjournalVolNumPages{Decision Support Systems}{85}{}{62--73}.
\PrintBackRefs{\CurrentBib}

\bibitem [\protect \citeauthoryear {%
Pedersen%
}{%
Pedersen%
}{%
{\protect \APACyear {2019}}%
}]{%
pedersen2019efficiently}
\APACinsertmetastar {%
pedersen2019efficiently}%
\begin{APACrefauthors}%
Pedersen, L\BPBI H.%
\end{APACrefauthors}%
\unskip\
\newblock
\APACrefYear{2019}.
\newblock
\APACrefbtitle {Efficiently inefficient: how smart money invests and market
  prices are determined} {Efficiently inefficient: how smart money invests and
  market prices are determined}.
\newblock
\APACaddressPublisher{}{Princeton University Press}.
\PrintBackRefs{\CurrentBib}

\bibitem [\protect \citeauthoryear {%
Rambaldi%
, Bacry%
\BCBL {}\ \BBA {} Muzy%
}{%
Rambaldi%
\ \protect \BOthers {.}}{%
{\protect \APACyear {2019}}%
}]{%
rambaldi2019disentangling}
\APACinsertmetastar {%
rambaldi2019disentangling}%
\begin{APACrefauthors}%
Rambaldi, M.%
, Bacry, E.%
\BCBL {}\ \BBA {} Muzy, J\BHBI F.%
\end{APACrefauthors}%
\unskip\
\newblock
\APACrefYearMonthDay{2019}{}{}.
\newblock
{\BBOQ}\APACrefatitle {Disentangling and quantifying market participant
  volatility contributions} {Disentangling and quantifying market participant
  volatility contributions}.{\BBCQ}
\newblock
\APACjournalVolNumPages{Quantitative Finance}{19}{10}{1613--1625}.
\PrintBackRefs{\CurrentBib}

\bibitem [\protect \citeauthoryear {%
Renault%
}{%
Renault%
}{%
{\protect \APACyear {2017}}%
}]{%
renault2017intraday}
\APACinsertmetastar {%
renault2017intraday}%
\begin{APACrefauthors}%
Renault, T.%
\end{APACrefauthors}%
\unskip\
\newblock
\APACrefYearMonthDay{2017}{}{}.
\newblock
{\BBOQ}\APACrefatitle {Intraday online investor sentiment and return patterns
  in the US stock market} {Intraday online investor sentiment and return
  patterns in the us stock market}.{\BBCQ}
\newblock
\APACjournalVolNumPages{Journal of Banking \& Finance}{84}{}{25--40}.
\PrintBackRefs{\CurrentBib}

\bibitem [\protect \citeauthoryear {%
Robert%
\ \BBA {} Rosenbaum%
}{%
Robert%
\ \BBA {} Rosenbaum%
}{%
{\protect \APACyear {2011}}%
}]{%
robert2011new}
\APACinsertmetastar {%
robert2011new}%
\begin{APACrefauthors}%
Robert, C\BPBI Y.%
\BCBT {}\ \BBA {} Rosenbaum, M.%
\end{APACrefauthors}%
\unskip\
\newblock
\APACrefYearMonthDay{2011}{}{}.
\newblock
{\BBOQ}\APACrefatitle {A new approach for the dynamics of ultra-high-frequency
  data: The model with uncertainty zones} {A new approach for the dynamics of
  ultra-high-frequency data: The model with uncertainty zones}.{\BBCQ}
\newblock
\APACjournalVolNumPages{Journal of Financial Econometrics}{9}{2}{344--366}.
\PrintBackRefs{\CurrentBib}

\bibitem [\protect \citeauthoryear {%
Robert%
\ \BBA {} Rosenbaum%
}{%
Robert%
\ \BBA {} Rosenbaum%
}{%
{\protect \APACyear {2012}}%
}]{%
robert2012volatility}
\APACinsertmetastar {%
robert2012volatility}%
\begin{APACrefauthors}%
Robert, C\BPBI Y.%
\BCBT {}\ \BBA {} Rosenbaum, M.%
\end{APACrefauthors}%
\unskip\
\newblock
\APACrefYearMonthDay{2012}{}{}.
\newblock
{\BBOQ}\APACrefatitle {Volatility and covariation estimation when
  microstructure noise and trading times are endogenous} {Volatility and
  covariation estimation when microstructure noise and trading times are
  endogenous}.{\BBCQ}
\newblock
\APACjournalVolNumPages{Mathematical Finance: An International Journal of
  Mathematics, Statistics and Financial Economics}{22}{1}{133--164}.
\PrintBackRefs{\CurrentBib}

\bibitem [\protect \citeauthoryear {%
Tetlock%
}{%
Tetlock%
}{%
{\protect \APACyear {2007}}%
}]{%
tetlock2007giving}
\APACinsertmetastar {%
tetlock2007giving}%
\begin{APACrefauthors}%
Tetlock, P\BPBI C.%
\end{APACrefauthors}%
\unskip\
\newblock
\APACrefYearMonthDay{2007}{}{}.
\newblock
{\BBOQ}\APACrefatitle {Giving content to investor sentiment: The role of media
  in the stock market} {Giving content to investor sentiment: The role of media
  in the stock market}.{\BBCQ}
\newblock
\APACjournalVolNumPages{The Journal of finance}{62}{3}{1139--1168}.
\PrintBackRefs{\CurrentBib}

\bibitem [\protect \citeauthoryear {%
Wyart%
, Bouchaud%
, Kockelkoren%
, Potters%
\BCBL {}\ \BBA {} Vettorazzo%
}{%
Wyart%
\ \protect \BOthers {.}}{%
{\protect \APACyear {2008}}%
}]{%
wyart2008relation}
\APACinsertmetastar {%
wyart2008relation}%
\begin{APACrefauthors}%
Wyart, M.%
, Bouchaud, J\BHBI P.%
, Kockelkoren, J.%
, Potters, M.%
\BCBL {}\ \BBA {} Vettorazzo, M.%
\end{APACrefauthors}%
\unskip\
\newblock
\APACrefYearMonthDay{2008}{}{}.
\newblock
{\BBOQ}\APACrefatitle {Relation between bid--ask spread, impact and volatility
  in order-driven markets} {Relation between bid--ask spread, impact and
  volatility in order-driven markets}.{\BBCQ}
\newblock
\APACjournalVolNumPages{Quantitative finance}{8}{1}{41--57}.
\PrintBackRefs{\CurrentBib}

\end{thebibliography}
    \bibliographystyle{apacite}
\end{document}